\documentclass[twocolumn,eqsecnum,aps]{revtex4}
\usepackage[dvips]{epsfig,color}
\usepackage{graphicx}
\usepackage{amsmath,amsfonts,amsbsy,amssymb}
\usepackage{tabularx}

\begin{document}

\title{ Supersymmetric Chern-Simons Theory and  Supersymmetric Quantum Hall Liquid}

\author{Kazuki Hasebe}
\affiliation{Department of General Education, Takuma National College of Technology,   Takuma-cho, Mitoyo-city, Kagawa 769-1192, Japan \\
Email: hasebe@dg.takuma-ct.ac.jp}

\begin{abstract}
 We develop a supersymmetric extension of  Chern-Simons theory and  
Chern-Simons-Landau-Ginzburg theory for the supersymmetric quantum Hall liquid.
Supersymmetric counterparts of topological and gauge  structures
  peculiar to the  Chern-Simons theory are inspected  in the supersymmetric Chern-Simons theory.
 We also explore an effective field theoretical description 
 for the supersymmetric quantum Hall liquid. 
 The key observation is the  the charge-flux duality.  Based on the duality, we derive a  dual
 supersymmetric Chern-Simons-Landau-Ginzburg  
 theory, and discuss  physical properties of the topological excitations 
 in the supersymmetric quantum Hall liquid.

\end{abstract}

\maketitle

\section{Introduction}

 In recent years, concepts of  quantum Hall effect (QHE), which was
 believed to be formulated only 
 in  two-dimensional space,
 have been dramatically changing,  initiated by  
  the success of the construction of the four-dimensional
 QHE \cite{cond-mat/0110572}.
 Since then,  further generalizations of the QHE have been explored, 
 in even higher dimensions 
 \cite{hep-th/0203264,cond-mat/0306045, hep-th/0309212, hep-th/0310274, hep-th/0505095} 
 and in $q$-deformed systems \cite{hep-th/0504092}.
More recently, inspired by  the developments of  the non-anticommutative geometry 
\cite{hep-th/0302109,%hep-th/0303063,
hep-th/0302078,hep-th/0305248}, 
   supersymmetric (SUSY)  extensions of  QH liquids
were proposed on noncommutative supermanifolds,  such as  
  a fuzzy supersphere \cite{hep-th/0411137} 
 and  a noncommutative superplane \cite{hep-th/0503162}.
 The analyses of the SUSY QHE  revealed some of the 
 novel physical consequences of the non-anticommutative geometry.
The SUSY QH liquids exhibit  natural SUSY counterparts of   
   the  mathematical and physical features peculiar to the original QH liquids,
 for instance,  noncommutative geometry, fractionally charged excitations, 
 Hall orthogonality and  $W_{\infty}$ symmetry. 
 The SUSY Landau problems on other analogous noncommutative supermanifolds, such as fuzzy $\mathbb{C}P^{(n|m)}$, 
 were studied in Ref.\cite{hep-th/0311159},  a higher SUSY Landau problem was also reported in Ref.\cite{hep-th/0503244}, and a SUSY extension of the QH matrix model was constructed in Ref.\cite{hep-th/0410070}.

One of the most amazing incidents of  QHE is the emergence of the 
 Chern-Simons (CS) topological field theory 
 in a low energy sector.
 The CS flux attachment  to electrons induces  
 the statistical transmutation from fermion to  (composite) boson,
 and the QH states are regarded as   ``superfluid'' states of  composite 
 bosons.
 The Chern-Simons-Landau-Ginzburg (CSLG) theory describes 
 low energy  phenomenology  in the QHE \cite{PRL8862,IJMP92B6,cond-mat/0206164}.
 Apart from relations to QHE, 
 the CS theory is important of its own right
 in a field theoretical point of view \cite{hep-th/9902115}.
 It is quite interesting  that such a novel
 field theory  appears from mysterious
  many-body 
 effects, and well describes collective phenomena of a real condensed matter system.

 In this paper, we develop a  SUSY extension of  the
CS theory and CSLG theory with $OSp(1|2)$ global symmetry.
 The SUSY CS theory
 demonstrates  natural SUSY extensions of the  topological and gauge 
 features peculiar to the  original CS theory.
  We also explore the CSLG description  for the 
 SUSY QH liquid. 
 We show the existence of the charge-flux duality in the SUSY QH system,
 and derive the 
 dual representation for the SUSY CSLG theory. 
 Based on the dual 
  description, physical properties of the
   topological excitations in   SUSY QH liquids are discussed. 

This paper is organized as follows. 
In Section \ref{Balachandranformalism},  we briefly review the mathematical background for the 
SUSY QH liquid.
In Section \ref{oneparticleLagrange}, we present a Lagrange formalism of the 
 one-particle mechanics on a supersphere in the presence of the supermonopole 
background.
In Section \ref{SUSYCSsec},   the SUSY CS theory is constructed
 and  its field theoretical properties are inspected.
In Section \ref{dualTPsec}, the charge-flux duality in the SUSY 
 system is explored.
 With the use of  the dual SUSY CSLG description, we discuss  physical properties of the topological excitations.
 Section \ref{summarysec} is devoted to summary and discussions.
In Appendix \ref{somformulurasupermatrix}, several useful formulas on supermatrix are summarized. 
In Appendix \ref{superjacobisec}, we present  the super Jacobi and Bianchi identities.  

%%%%%%%%%%%%%%%%%%%%%%%%%%%%%%%%%%%%%%%%%%%%%%%%%%%%%%%%%%
\section{The super Hopf map and the supermonopole}\label{Balachandranformalism}
%%%%%%%%%%%%%%%%%%%%%%%%%%%%%%%%%%%%%%%%%%%%%%%%%%%%%%%%

In this section, we review the super Hopf map and 
 the supermonopole \cite{math-ph/9907020,hep-th/0409230} used in the set-up of the supersymmetric QH liquid.
First, we  introduce the $OSp(1|2)$ supergroup, whose generators  
 are given by  $l_a$  $(a=x,y,z)$, which are Grassmann even, and $l_{\alpha}$ $(\alpha=\theta_1,\theta_2)$, which are Grassmann odd.
 They satisfy the following graded algebras:
%%%%%%%%%%%%%%%%%%%%%%%%%%%%%%%%%%%%%%%%%%%%%%%%%%%%%%%%%%%%%%
\begin{subequations}
\begin{align}
&[l_a,l_b]=i\epsilon_{abc}l_c,\\
&[l_a,l_{\alpha}]=\frac{1}{2}(\sigma_{a})_{\beta\alpha}l_{\beta},\\
&\{l_{\alpha},l_{\beta}\}=\frac{1}{2}(C\sigma_{a})_{\alpha\beta}l_a.
\end{align}
\end{subequations}
%%%%%%%%%%%%%%%%%%%%%%%%%%%%%%%%%%%%%%%%%%%%%%%%%%%%%%%%%%%%%%%
The fundamental representations of the $OSp(1|2)$ generators are given by 
%%%%%%%%%%%%%%%%%%%%%%%%%%%%%%%%%%%%%%%%%%%%%%%%%%%%%%
\begin{equation}
l_a=\frac{1}{2}
\begin{pmatrix}
\sigma_a & 0 \\
 0 & 0 
\end{pmatrix},~~~
l_{\alpha} =i\frac{1}{2}
\begin{pmatrix}
0 & \tau_{\alpha} \\
(C\tau_{\alpha})^t & 0
\end{pmatrix},
\label{fundamentalgene}
\end{equation}
%%%%%%%%%%%%%%%%%%%%%%%%%%%%%%%%%%%%%%%%%%%%%%%%%%%%%%
where $\{\sigma_a\}$ are Pauli matrices, 
$C$ is the charge conjugation matrix given by $C=i\sigma_2$ and $\tau_1=(1,0)^t$, $\tau_2=(0,1)^t$.
The general element of the $OSp(1|2)$ supergroup 
(strictly speaking $UOSp(1|2)$ supergroup), is parameterized as 
%%%%%%%%%%%%%%%%%%%%%%%%%%%%%%%%%%%%%%%%%%%%%%%%%%%%%%%%%
\begin{equation}
g=
\begin{pmatrix}
& u & -v^{*} & \eta^* u- \eta v^* \\ 
& v & u^* & \eta u^* +\eta^* v \\
& \eta & -\eta^* & 1- \eta^*\eta 
\end{pmatrix},
\end{equation}
%%%%%%%%%%%%%%%%%%%%%%%%%%%%%%%%%%%%%%%%%%%%%%%%%%%%%%%%%
where $u$ and $v$ are Grassmann even complex parameters,  $\eta$ is a Grassmann odd parameter, and 
 they are chosen to 
 satisfy the constraint $u^* u+v^*v -\eta^*\eta=1$.
With this constraint, one may see that 
 $g$ satisfies the following conditions
%%%%%%%%%%%%%%%%%%%%%%%%%%%%%%%%%%%%%%%%%%%%%%%%%%%%%%%%%
\begin{equation}
sdet (g)= u^* u+v^*v -\eta^*\eta=1
\end{equation}
%%%%%%%%%%%%%%%%%%%%%%%%%%%%%%%%%%%%%%%%%%%%%%%%%%%%%%%%%%%
and 
%%%%%%%%%%%%%%%%%%%%%%%%%%%%%%%%%%%%%%%%%%%%%%%%%%%%%%%%%%%%
\begin{equation}
g^{\ddagger}g=g g^{\ddagger}=1.
\end{equation}
%%%%%%%%%%%%%%%%%%%%%%%%%%%%%%%%%%%%%%%%%%%%%%%%%%%%%%%%%%%
The definitions of the super-determinant $sdet$,
 and the super-adjoint ${\ddagger}$ are given by  Eqs.(\ref{superdeterminantformula}) and (\ref{super-adjointformula}), respectively. (Our definition of the super-adjoint is 
different from the conventional one.)

The super Hopf map is given by the mapping from  $OSp(1|2)$ element $g$ to
 the coordinates $(x_a,\theta_{\alpha})$ on the supersphere $S^{2|2}$
%%%%%%%%%%%%%%%%%%%%%%%%%%%%%%%%%%%%%%%%%%%%%%%%%%%%%%%%%
\begin{equation}
g \rightarrow g l_3 g^{\ddagger}= x_a l_a + iC_{\alpha\beta}\theta_{\alpha} l_{\beta}.
\label{gHopfmapping}
\end{equation}
%%%%%%%%%%%%%%%%%%%%%%%%%%%%%%%%%%%%%%%%%%%%%%%%%%%%%%%%
Taking  square and supertrace on both sides (See Eqs.(\ref{supertraceformula}) and
 (\ref{supertracefund})), it is easily checked  that  $(x_a,\theta_{\alpha})$ satisfy the constraint 
%%%%%%%%%%%%%%%%%%%%%%%%%%%%%%%%%%%%%%%%%%%%%%%%%%%%%%%%%%%%%%
\begin{equation}
x_a^2+C_{\alpha\beta}\theta_{\alpha}\theta_{\beta}=1,
\label{supersphereconstraint}
\end{equation}
%%%%%%%%%%%%%%%%%%%%%%%%%%%%%%%%%%%%%%%%%%%%%%%%%%%%%%%%%%%
which defines the supersphere with unit radius.
With the use of  the Hopf spinor $\psi=(u,v,\eta)^t$,  
  $(x_a,\theta_{\alpha})$ in Eq.(\ref{gHopfmapping}) 
 are concisely represented as 
%%%%%%%%%%%%%%%%%%%%%%%%%%%%%%%%%%%%%%%%%%%%%%%%%%%%%%%%%%
\begin{subequations}
\begin{align}
&x_a=2\psi^{\dagger} l_a\psi,\\
&\theta_{\alpha}=-2i\psi^{\dagger}l_{\alpha}\psi,
\end{align}\label{1stHopf2}
\end{subequations}
%%%%%%%%%%%%%%%%%%%%%%%%%%%%%%%%%%%%%%%%%%%%%%%%%%%%%%%%%
where $\psi^{\dagger}\equiv (u^*,v^*,\eta^*)$. 
Since $(x_a, \theta_{\alpha})$ are  invariant under the $U(1)$ transformation
%%%%%%%%%%%%%%%%%%%%%%%%%%%%%%%%%%%%%%%%%%%%%%%%%%%%%%%%%
\begin{equation}
g\rightarrow g\cdot e^{2i\alpha l_3}
\label{gsuperU1}
\end{equation}
%%%%%%%%%%%%%%%%%%%%%%%%%%%%%%%%%%%%%%%%%%%%%%%%%%%%%%%%%%  
or
%%%%%%%%%%%%%%%%%%%%%%%%%%%%%%%%%%%%%%%%%%%%%%%%%%%%%%%%%
\begin{equation}
\psi\rightarrow e^{i\alpha}\psi,
\label{gsuperU2}
\end{equation}
%%%%%%%%%%%%%%%%%%%%%%%%%%%%%%%%%%%%%%%%%%%%%%%%%%%%%%%%%
the supersphere is given by the coset $S^{2|2}=OSp(1|2)/U(1)$, and    
   $U(1)$ fibre is  defined on $S^{2|2}$.
The connection of the $U(1)$ fiber is given by  
%%%%%%%%%%%%%%%%%%%%%%%%%%%%%%%%%%%%%%%%%%%%%%%%%%%%%%%%%%%%
\begin{equation}
A =-{i} str (l_3 g^{\ddagger}dg)=-i\frac{1}{2}(\psi^{\ddagger}d\psi - d\psi^{\ddagger}\psi),
\end{equation}
%%%%%%%%%%%%%%%%%%%%%%%%%%%%%%%%%%%%%%%%%%%%%%%%%%%%%%%%%%%
where $\psi^{\ddagger}\equiv (u^*,v^*,-\eta^*)$. 
Under the $U(1)$ transformation (\ref{gsuperU1}) or (\ref{gsuperU2}),  as expected,
 $A$  is transformed as  
%%%%%%%%%%%%%%%%%%%%%%%%%%%%%%%%%%%%%%%%%%%%%%%%%%%%%%%%
\begin{equation} 
A\rightarrow A+d\alpha.
\end{equation}
%%%%%%%%%%%%%%%%%%%%%%%%%%%%%%%%%%%%%%%%%%%%%%%%%%%%%%%

Inverting  the super Hopf map (\ref{1stHopf2}) from $(x_a,\theta_{\alpha})$ to 
$\psi$, the super Hopf spinor is expressed as 
\cite{hep-th/0409230},
%%%%%%%%%%%%%%%%%%%%%%%%%%%%%%%%%%%%%%%%%%%%%%%%%%%%%%%%%%
\begin{equation}
\psi=
\begin{pmatrix}
&\sqrt{\frac{1+x_3}{2}}\biggl(1-\frac{1}{4(1+x_3)}\theta C\theta\biggr)\\
&\frac{x_1+ix_2}{\sqrt{2(1+x_3)}}\biggl(1+\frac{1}{4(1+x_3)}\theta C\theta \biggr)\\
&\frac{1}{\sqrt{2(1+x_3)}}\biggl((1+x_3)\theta_1+(x_1+ix_2)\theta_2\biggr)
\end{pmatrix}.
\label{superhopfspi}
\end{equation}
%%%%%%%%%%%%%%%%%%%%%%%%%%%%%%%%%%%%%%%%%%%%%%%%%%%%%%%%%%%%%
Using this explicit form, 
 the supermonopole gauge fields $A= dx_a A_a +d\theta_{\alpha} A_{\alpha}$ are
 calculated as 
%%%%%%%%%%%%%%%%%%%%%%%%%%%%%%%%%%%%%%%%%%%%%%%%%%%%%%
\begin{subequations}
\begin{align}
&A_a= -\frac{I}{2}\epsilon_{ab3}\frac{x_b}{1+x_3}\biggl( 1+\frac{2+x_3}{2(1+x_3)}
\theta C\theta   \biggr),\\
&A_{\alpha}=-i\frac{I}{2}(\sigma_a C)_{\alpha\beta}x_a\theta_{\beta},
\end{align}\label{supermonopolegauge}
\end{subequations}
%%%%%%%%%%%%%%%%%%%%%%%%%%%%%%%%%%%%%%%%%%%%%%%%%%%%%%%%%
with $I=1$. 
The supermonopole gauge fields with quantized charges 
take the same form as Eqs.(\ref{supermonopolegauge}) with the integer $I$. 
With the use of super gauge fields (\ref{supermonopolegauge}), the supermonopole field strengths  
 are obtained as 
%%%%%%%%%%%%%%%%%%%%%%%%%%%%%%%%%%%%%%%%%%%%%%%%%%%%%
\begin{subequations}
\begin{align}
&F_{ab}=\frac{I}{2}\epsilon_{abc}x_c\biggl(1+\frac{3}{2}\theta C\theta\biggr),\\
&F_{a\alpha}=-i\frac{I}{2}(\delta_{ab}-3{x_a x_b} ) (\theta \sigma_b C)_{\alpha},\\
&F_{\alpha\beta}=-iI(\sigma_a C)_{\alpha\beta}x_a\biggl(1+\frac{3}{2}\theta C\theta\biggr),
\end{align}\label{supermonopolefieldst}
\end{subequations}
%%%%%%%%%%%%%%%%%%%%%%%%%%%%%%%%%%%%%%%%%%%%%%%%%%%%%%%%%%%%%%%%%%%%%%%%%%%%%%%%%%%%%%%%%%%%%%%%%
where we used the definition of the super field strengths  
%%%%%%%%%%%%%%%%%%%%%%%%%%%%%%%%%%%%%%%%%%%%%%%%%%%%%
\begin{subequations}
\begin{align}
&F_{ab}=\partial_a A_b-\partial_b A_a,\\
&F_{a\alpha}=\partial_{a}A_{\alpha}-\partial_{\alpha}A_a,\\
&F_{\alpha\beta}=\partial_{\alpha}A_{\beta}+\partial_{\beta}A_{\alpha}.
\end{align}
\end{subequations}
%%%%%%%%%%%%%%%%%%%%%%%%%%%%%%%%%%%%%%%%%%%%%%%%%%%%%%%%%%%%%%%%%%%%%%%%%%%%%%%%%%%%%%%%%%%%%%%%%

%%%%%%%%%%%%%%%%%%%%%%%%%%%%%%%%%%%%%%%%%%%%%%%%%%%%%%%%
\section{one-particle on the supersphere in the supermonopole background}\label{oneparticleLagrange}
%%%%%%%%%%%%%%%%%%%%%%%%%%%%%%%%%%%%%%%%%%%%%%%%%%%%%%%%

Before discussing the many-body system, we consider   
one-particle mechanics on the supersphere with unit radius in the 
   supermonopole background. 
The supermonopole is set at the center of the supersphere. 
The  one-particle Lagrangian is given by 
%%%%%%%%%%%%%%%%%%%%%%%%%%%%%%%%%%%%%%%%%%%%%%%%%%%%%%%%%
\begin{equation}
L=\frac{m}{2}({\dot{x}_a}^2+
C_{\alpha\beta}\dot{\theta_{\alpha}}\dot{\theta_{\beta}})
+\dot{x}_a A_a +\dot{\theta}_{\alpha} A_{\alpha}-V,
\label{oneparticelonsupersphere}
\end{equation}
%%%%%%%%%%%%%%%%%%%%%%%%%%%%%%%%%%%%%%%%%%%%%%%%%%%%%%%%%
where $V$ is the external electric potential, $(A_a,A_{\alpha})$ are 
 supermonopole gauge fields (\ref{supermonopolegauge}),
 and $(x_a,\theta_{\alpha})$ satisfy the constraint (\ref{supersphereconstraint}).
Introducing a Lagrange multiplier $\lambda$, 
the equations of motion are derived as 
%%%%%%%%%%%%%%%%%%%%%%%%%%%%%%%%%%%%%%%%%%%%%%%%%%%%%%%%%
\begin{subequations}
\begin{align}
&m\ddot{x}_a=F_{ab}\dot{x}_b-F_{a\alpha}\dot{\theta}_{\alpha}+E_a+\lambda x_a,\\ 
&m\ddot{\theta}_{\alpha}=C_{\alpha\beta}( F_{a\beta}\dot{x}_a
+F_{\beta\gamma}\dot{\theta}_{\gamma})+E_{\alpha}+\lambda\theta_{\alpha},
\end{align}\label{EOMCSparticle}
\end{subequations}
%%%%%%%%%%%%%%%%%%%%%%%%%%%%%%%%%%%%%%%%%%%%%%%%%%%%%%%%
where $E_a=-\partial_a V$,  
$E_{\alpha}=C_{\alpha\beta}\partial_{\beta} V$, and 
$(F_{ab},F_{a\alpha},F_{\alpha\beta})$ are supermonopole gauge fields (\ref{supermonopolefieldst}).
From Eqs.(\ref{supersphereconstraint}) and (\ref{EOMCSparticle}),
 $\lambda$ is  eliminated as 
%%%%%%%%%%%%%%%%%%%%%%%%%%%%%%%%%%%%%%%%%%%%%%%%%%%%%%%%%%%%%%%%
\begin{equation}
\lambda=-{m}({\dot{x}}_a^2+C_{\alpha\beta}\dot{\theta}_{\alpha}\dot{\theta}_{\beta})
-(E_a x_a+C_{\alpha\beta}E_{\alpha}\theta_{\beta}).
\end{equation}
%%%%%%%%%%%%%%%%%%%%%%%%%%%%%%%%%%%%%%%%%%%%%%%%%%%%%%%%%%%%%%%%
Inserting this explicit form of $\lambda$ to Eqs.(\ref{EOMCSparticle}), 
  we obtain selfcontained  equations of motion.
 Though it is quite nontrivial  to solve such nonlinear equations,  
  we may discuss local motions of the particle of our interest.

The center-of-mass coordinates $(X_a,\Theta_{\alpha})$ are defined as 
%%%%%%%%%%%%%%%%%%%%%%%%%%%%%%%%%%%%%%%%%%%%%%%%%%%%%%%%%%%%
\begin{subequations}
\begin{align}
&X_a=x_a-\frac{2}{I} \Lambda_a,\\
&\Theta_{\alpha}=\theta_{\alpha}-\frac{2}{I}\Lambda_{\alpha},
\end{align}\label{coordinatescenter}
\end{subequations}
%%%%%%%%%%%%%%%%%%%%%%%%%%%%%%%%%%%%%%%%%%%%%%%%%%%%%%%%%
where $(\Lambda_a,\Lambda_{\alpha})$ represent the $OSp(1|2)$ angular momenta
of the particle 
%%%%%%%%%%%%%%%%%%%%%%%%%%%%%%%%%%%%%%%%%%%%%%%%%%%%%%%%%
\begin{subequations}
\begin{align}
&\Lambda_a={m}\epsilon_{abc} x_b \dot{x}_c+ 
i\frac{m}{2}\theta_{\alpha}(\sigma_a C)_{\alpha\beta}\dot{\theta}_{\beta},\\
&\Lambda_{\alpha}=i\frac{m}{2}x_a(\sigma_a)_{\beta\alpha}\dot{\theta}_{\beta}
-i\frac{m}{2}\theta_{\beta}(\sigma_a)_{\beta\alpha}\dot{x}_a.
\end{align}
\end{subequations}
%%%%%%%%%%%%%%%%%%%%%%%%%%%%%%%%%%%%%%%%%%%%%%%%%%%%%%%%%
In the lowest Landau level (LLL) limit (which is realized at  $m \rightarrow 0$),
 the  particle coordinates $(x_a,\theta_{\alpha})$ are reduced to the center-of-mass coordinates $(X_a,\Theta_{\alpha})$.
 With the use of  Eqs.(\ref{EOMCSparticle}), the constraint 
(\ref{supersphereconstraint}), and its derivative $\dot{x}_a x_a+C_{\alpha\beta}\dot{\theta}_{\alpha}\theta_{\beta}=0$,  
 the velocities of the center-of-mass coordinates are derived as 
%%%%%%%%%%%%%%%%%%%%%%%%%%%%%%%%%%%%%%%%%%%%%%%%%%%%%%%%%%%%%%
\begin{subequations}
\begin{align}
&\dot{X}_a=-\frac{I}{2}\epsilon_{abc}x_b E_c-i\frac{I}{4}\theta_{\alpha}
(\sigma_a C)_{\alpha\beta}E_{\beta},\\
&\dot{\Theta}_{\alpha}=-i\frac{I}{4}(\sigma_a)_{\beta\alpha}x_a E_{\beta}+i\frac{I}{4}
(\sigma_a)_{\beta\alpha}\theta_{\beta}E_a.
\end{align}\label{velocityceonter}
\end{subequations}
%%%%%%%%%%%%%%%%%%%%%%%%%%%%%%%%%%%%%%%%%%%%%%%%%%%%%%%%%%%%%
In the presence of the magnetic field, a charged
particle performs  a drift motion, where 
 the center-of-mass coordinates  move perpendicularly to the direction of the 
applied electric fields. 
From  Eqs.(\ref{velocityceonter}), we confirm such orthogonality in the SUSY sense:  
 %%%%%%%%%%%%%%%%%%%%%%%%%%%%%%%%%%%%%%%%%%%%%%%%%%%%%%%%%%%%%
\begin{equation}
E_a \dot{X}_a+C_{\alpha\beta}E_{\alpha}\dot{\Theta}_{\beta}=0.
\end{equation}
%%%%%%%%%%%%%%%%%%%%%%%%%%%%%%%%%%%%%%%%%%%%%%%%%%%%%%%%%%%%%
Meanwhile, from Eqs.(\ref{EOMCSparticle}), the  particle velocities 
and the electric fields are related as  
%%%%%%%%%%%%%%%%%%%%%%%%%%%%%%%%%%%%%%%%%%%%%%%%%%%%%%%%%%%%%
\begin{equation}
E_a \dot{x}_a  +C_{\alpha\beta} E_{\alpha} \dot{\theta}_{\beta}=
m(\dot{x}_a \ddot{x}_a+C_{\alpha\beta}\dot{\theta}_{\alpha}\ddot{\theta}_{\beta}).
\label{relationvelacc}
\end{equation}
%%%%%%%%%%%%%%%%%%%%%%%%%%%%%%%%%%%%%%%%%%%%%%%%%%%%%%%%%%%%%
In the LLL limit, from Eq.(\ref{relationvelacc}), 
 one may find the super Hall orthogonality 
%%%%%%%%%%%%%%%%%%%%%%%%%%%%%%%%%%%%%%%%%%%%%%%%%%%%%%%%%%
\begin{equation}
E_a \dot{x}_a+C_{\alpha\beta}E_{\alpha}\dot{\theta}_{\beta}=0,
\end{equation}
%%%%%%%%%%%%%%%%%%%%%%%%%%%%%%%%%%%%%%%%%%%%%%%%%%%%%%%%%%
which is consistent with the one   
obtained in the SUSY noncommutative formalism \cite{hep-th/0411137}.

Without electric fields, 
from Eqs.(\ref{velocityceonter}),  the velocities of the center-of-mass coordinates  vanish 
%%%%%%%%%%%%%%%%%%%%%%%%%%%%%%%%%%%%%%%%%%%%%%%%%%%%%%%%%%%%%%%%%%%%%%%%%%
\begin{equation}
\dot{X}_a=\dot{\Theta}_{\alpha}=0,
\label{zerovolcenter}
\end{equation}
%%%%%%%%%%%%%%%%%%%%%%%%%%%%%%%%%%%%%%%%%%%%%%%%%%%%%%%%%%%%%%%%%%%%%%%%%%%
and, from Eq.(\ref{relationvelacc}),  
the accelerations and velocities become orthogonal
%%%%%%%%%%%%%%%%%%%%%%%%%%%%%%%%%%%%%%%%%%%%%%%%%%%%%%%%%%%
\begin{equation}
\dot{x}_a\ddot{x}_a 
+C_{\alpha\beta}\dot{\theta}_{\alpha} \ddot{\theta}_{\beta} =0.
\label{orthonalvelacc}
\end{equation}
%%%%%%%%%%%%%%%%%%%%%%%%%%%%%%%%%%%%%%%%%%%%%%%%%%%%%%%%%%%%%%
 Under the perpendicular magnetic field, 
 a charged particle  performs a cyclotron motion around its center-of-mass coordinates due to the Lorentz force.
Equations (\ref{zerovolcenter}) and (\ref{orthonalvelacc}) demonstrate this observation  in the SUSY sense.

In the planar limit $(x_3\approx 1)$, the one-particle Lagrangian (\ref{oneparticelonsupersphere}) is reduced to 
%%%%%%%%%%%%%%%%%%%%%%%%%%%%%%%%%%%%%%%%%%%%%%%%%%%%%%%%%%%
\begin{equation}
L=\frac{m}{2}(\dot{x}_i^2+C_{\alpha\beta}\dot{\theta}_{\alpha}\dot{\theta}_{\beta})-
\frac{B}{2}\epsilon_{ij}\dot{x}_i x_j-iB(\sigma_1)_{\alpha\beta}\dot{\theta}_{\alpha}\theta_{\beta},
\label{planarsusylag}
\end{equation}
%%%%%%%%%%%%%%%%%%%%%%%%%%%%%%%%%%%%%%%%%%%%%%%%%%%%%%%%%%%%
where $B=I/2$.
The canonical 
momenta are obtained as 
%%%%%%%%%%%%%%%%%%%%%%%%%%%%%%%%%%%%%%%%%%%%%%%%%%%%%%%%%%%%%%%
\begin{subequations}
\begin{align}
&p_i=\frac{\partial}{\partial \dot{x}_i}L=m\dot{x}_i-\frac{B}{2}\epsilon_{ij}x_j,\\
&p_{\alpha}=\frac{\partial}{\partial\dot{\theta}_{\alpha}}L=mC_{\alpha\beta}\dot{\theta}_{\beta}-iB(\sigma_1)_{\alpha\beta}\theta_{\beta}.
\end{align}
\end{subequations}
%%%%%%%%%%%%%%%%%%%%%%%%%%%%%%%%%%%%%%%%%%%%%%%%%%%%%%%%%%%%%%
In the LLL limit, from the 
commutation relations $[x_i,p_j]=i\delta_{ij}$ and $\{\theta_{\alpha},p_{\beta}\}=i\delta_{\alpha\beta}$,
 we obtain the
algebras on the noncommutative superplane
%%%%%%%%%%%%%%%%%%%%%%%%%%%%%%%%%%%%%%%%%%%%%%%%%%%%%%%%%%%%%%%%%%
\begin{subequations}
\begin{align}  
&[x_i,x_j]=-i\frac{1}{B}\epsilon_{ij},\\
&\{\theta_{\alpha},\theta_{\beta}\}=-\frac{1}{2B} (\sigma_1)_{\alpha\beta}.
\end{align}\label{ncalgebrasuperplane}
\end{subequations} 
%%%%%%%%%%%%%%%%%%%%%%%%%%%%%%%%%%%%%%%%%%%%%%%%%%%%%%%%%%%%%%%%%%%%%%%
These SUSY noncommutative algebras bring novel physics to  
planar SUSY 
quantum  Hall 
 systems \cite{hep-th/0503162}.

%%%%%%%%%%%%%%%%%%%%%%%%%%%%%%%%%%%%%%%%%%%%%%%%%%%%%%%%%
\section{SUSY CS theory}\label{SUSYCSsec}
%%%%%%%%%%%%%%%%%%%%%%%%%%%%%%%%%%%%%%%%%%%%%%%%%%%%%%%%%%

It is well known that CS theories are defined in spaces
 with odd dimensions.
With the CS coupling constant $\kappa$, the   CS Lagrangian in 3-dimensional space 
is given by  
%%%%%%%%%%%%%%%%%%%%%%%%%%%%%%%%%%%%%%%%%%%%%%%%%%%%%%%%%%%%%%%%%%
\begin{equation}
\mathcal{L}_{CS}=\frac{\kappa}{4}\epsilon_{abc}A_a F_{bc}.
\label{originalCS}
\end{equation}
%%%%%%%%%%%%%%%%%%%%%%%%%%%%%%%%%%%%%%%%%%%%%%%%%%%%%%%%%%%%%%%%%%%
In the following, we modify $\mathcal{L}_{CS}$ to be invariant under the $OSp(1|2)$ global supersymmetry. 

 As a preliminary, we summarize the 
 $OSp(1|2)$ transformations of the super gauge fields.
The derivative expressions for the $OSp(1|2)$ generators are 
%%%%%%%%%%%%%%%%%%%%%%%%%%%%%%%%%%%%%%%%%%%%%%%%%%%%%%%%%
\begin{subequations}
\begin{align}
&L_a=-i\epsilon_{abc}x_b\partial_c
+\frac{1}{2}\theta_{\alpha}(\sigma_a)_{\alpha\beta}\partial_{\beta},\\
&L_{\alpha}=\frac{1}{2}x_a(C\sigma_a)_{\alpha\beta}\partial_{\beta}
-\frac{1}{2}\theta_{\beta}(\sigma_a)_{\beta\alpha}\partial_a.
\end{align}
\end{subequations}
%%%%%%%%%%%%%%%%%%%%%%%%%%%%%%%%%%%%%%%%%%%%%%%%%%%%%%%%%%
With the use of the  Grassmann odd generators $L_{\alpha}$,
  the super charge is constructed as   
%%%%%%%%%%%%%%%%%%%%%%%%%%%%%%%%%%%%%%%%%%%%%%%%%%%%%%%%%%%%%%
\begin{equation}
Q=L_{\alpha}\xi_{\alpha},
\end{equation}
%%%%%%%%%%%%%%%%%%%%%%%%%%%%%%%%%%%%%%%%%%%%%%%%%%%%%%%%%%%%%
where $\xi_{\alpha}$ are  Grassmann odd parameters.
The super transformations of  $(x_a,\theta_{\alpha})$ read
%%%%%%%%%%%%%%%%%%%%%%%%%%%%%%%%%%%%%%%%%%%%%%%%%%%%%%%
\begin{subequations}
\begin{align}
&\delta_{\xi}x_a=[Q,x_a]=\frac{1}{2}\theta\sigma_a \xi,\\
&\delta_{\xi}\theta_{\alpha}=\frac{1}{2}x_a(C\sigma_a\xi)_{\alpha}.
\end{align}
\end{subequations}
%%%%%%%%%%%%%%%%%%%%%%%%%%%%%%%%%%%%%%%%%%%%%%%%%%%%%%%
Similarly, the super gauge fields and the super field strengths are transformed as 
%%%%%%%%%%%%%%%%%%%%%%%%%%%%%%%%%%%%%%%%%%%%%%%%%%%%%%%%%%%%%
\begin{subequations}
\begin{align}
&\delta_{\xi}A_a=-\frac{1}{2}A_{\alpha}(C\sigma_a)_{\alpha\beta}\xi_{\beta},\\
&\delta_{\xi}A_{\alpha}=\frac{1}{2}A_a(\sigma_a\xi)_{\alpha},
\end{align}\label{SUSYAgauge}
\end{subequations}
%%%%%%%%%%%%%%%%%%%%%%%%%%%%%%%%%%%%%%%%%%%%%%%%%%%%%%%%%%%%
and 
%%%%%%%%%%%%%%%%%%%%%%%%%%%%%%%%%%%%%%%%%%%%%%%%%%%%%%%%%%
\begin{subequations}
\begin{align}
&\delta_{\xi}F_{ab}= -\frac{1}{2}F_{a\alpha}(C\sigma_b \xi)_{\alpha}+\frac{1}{2}F_{b\alpha}(C\sigma_a \xi)_{\alpha},\\
&\delta_{\xi}F_{a\alpha}=\frac{1}{2}F_{ab}(\sigma_b\xi)_{\alpha}
+\frac{1}{2}F_{\alpha\beta}(C\sigma_a \xi)_{\beta},\\
&\delta_{\xi}F_{\alpha\beta}= -\frac{1}{2}F_{a\alpha}(\sigma_a\xi)_{\beta}
-\frac{1}{2}F_{a\beta}(\sigma_a\xi)_{\alpha}.
\end{align}\label{SUSYFstren}
\end{subequations}
%%%%%%%%%%%%%%%%%%%%%%%%%%%%%%%%%%%%%%%%%%%%%%%%%%%%%%%%%%%%
It is noted that  $\bold{12}$-dimensional representations $(F_{ab},F_{a\alpha},F_{\alpha\beta})$ 
are $\it{not}$ irreducible representations 
 but  irreducibly decomposed into $\bold{7}\oplus\bold{5}$ representations.
The $\bold{5}$-dimensional irreducible representations $(F_a,F_{\alpha})$, 
which we call the super vector field strengths, are constructed as  
%%%%%%%%%%%%%%%%%%%%%%%%%%%%%%%%%%%%%%%%%%%%%%%%%%%%%%%%%%%%%%%%%%%%%%%%%%%%%%%%%%%%%%%%%%
\begin{subequations}
\begin{align}
&F_a=\frac{1}{2}\epsilon_{abc}F_{bc}+i\frac{1}{4}(C\sigma_a)_{\alpha\beta}F_{\alpha\beta},\\
&F_{\alpha}=-i\frac{1}{2}(C\sigma_a)_{\alpha\beta}F_{a\beta}.
\end{align}\label{bold5dim}
\end{subequations}
%%%%%%%%%%%%%%%%%%%%%%%%%%%%%%%%%%%%%%%%%%%%%%%%%%%%%%%%%%%%%%%%%%%%%%%%%%%%%%%%%%%%%%%%%%    
It is easy to check that, under the super transformations (\ref{SUSYFstren}), 
 $(F_a,F_{\alpha})$ form a super multiplet 
%%%%%%%%%%%%%%%%%%%%%%%%%%%%%%%%%%%%%%%%%%%%%%%%%%%%%%%%%%%%%
\begin{subequations}
\begin{align}
&\delta_{\xi}F_a=\frac{1}{2}F_{\alpha}(\sigma_a \xi)_{\alpha}, \\
&\delta_{\xi}F_{\alpha}= \frac{1}{2}F_a (C\sigma_a\xi)_{\alpha},
\label{SUSY5F}
\end{align}
\end{subequations}
%%%%%%%%%%%%%%%%%%%%%%%%%%%%%%%%%%%%%%%%%%%%%%%%%%%%%%%%%%%%
and satisfy 
  the scalar super  Bianchi identity
%%%%%%%%%%%%%%%%%%%%%%%%%%%%%%%%%%%%%%%%%%%%%%%%%%%%%%%%%%%%%%%%%%%%%%%%%%%%%%%%%%%
\begin{equation}
\partial_a F_a +\partial_{\alpha} F_{\alpha}=0,
\label{superBianchiire}
\end{equation}
%%%%%%%%%%%%%%%%%%%%%%%%%%%%%%%%%%%%%%%%%%%%%%%%%%%%%%%%%%%%%%%%%%%%%%%%%%%%%%%%%%%%%%% 
where we used the identities
$\epsilon_{abc}\partial_a F_{bc}=0$ and $(C\sigma_a)_{\alpha\beta}(2\partial_{\alpha}F_{\beta a}-\partial_{a}F_{\alpha\beta})=0$, which are obtained from the  
super Bianchi identities (\ref{superb1}) and (\ref{superb3}).

From the inner product of $(A_a,A_{\alpha})$ and $(F_a,F_{\alpha})$,  an  $OSp(1|2)$ singlet is constructed  as 
%%%%%%%%%%%%%%%%%%%%%%%%%%%%%%%%%%%%%%%%%%%%%%%%%%%%%%%%%%%%%%%%%%%%%%%%%%%%%%%%%%%%%%%%%%%%
\begin{equation}
\mathcal{L}_{sCS}= \frac{\kappa}{2}(A_a F_a +A_{\alpha}F_{\alpha}),
\label{SUSYCS5}
\end{equation} 
%%%%%%%%%%%%%%%%%%%%%%%%%%%%%%%%%%%%%%%%%%%%%%%%%%%%%%%%%%%%%%%%%%%%%%%%%%%%%%%%%%%%%%%%%%%
which we adopt as the  SUSY CS Lagrangian.
In terms of $(F_{ab},F_{a\alpha},F_{\alpha\beta})$, 
$\mathcal{L}_{sCS}$ is rewritten  as 
%%%%%%%%%%%%%%%%%%%%%%%%%%%%%%%%%%%%%%%%%%%%%%%%%%%%%%%%%%
\begin{align}
&\mathcal{L}_{sCS}\nonumber\\
&= \frac{\kappa}{4}(\epsilon_{abc}A_aF_{bc}-i(C\sigma_a)_{\alpha\beta}A_{\alpha}F_{a\beta}
+\frac{i}{2}(C\sigma_a)_{\alpha\beta}A_aF_{\alpha\beta})\nonumber\\
&=\frac{\kappa}{2}(\epsilon_{abc}A_a\partial_b A_c-\frac{i}{2}(C\sigma_a)_{\alpha\beta}A_{\alpha}\partial_a A_{\beta}+i(C\sigma_a)_{\alpha\beta}A_{\alpha}\partial_{\beta}A_a) \nonumber\\
&~~~ +(\text{total fermionic derivative term}).
\label{SUSYCS}
\end{align}
%%%%%%%%%%%%%%%%%%%%%%%%%%%%%%%%%%%%%%%%%%%%%%%%%%%%%%%%%%
Though either term in $\mathcal{L}_{sCS}$ is not $OSp(1|2)$ singlet, 
$\mathcal{L}_{sCS}$ is  
invariant in total. To respect the $OSp(1|2)$ global symmetry, 
the basespace for the SUSY CS Lagrangian is given by $\mathbb{R}^{3|2}$, whose 
 volume element is   $d^3x d^2\theta$.
 Thus, we have obtained the SUSY CS action  invariant under the $OSp(1|2)$ super transformation,
 while it is   $\it{not}$ invariant under general super
 coordinate transformations.
 Namely, the form of our SUSY CS action critically depends on the particular 
 choice of the coordinates and the  background manifold. 
 Then,  in this sense,  
 our SUSY CS theory is $\it{not}$ a topological field theory on supermanifolds.
(Topological field theories do not depend on the background metric and are invariant under 
 general coordinate transformations.)
%(The topological field theoretical  feature  may  only be 
%  referred to its
%bosonic counterpart which appears as the first term in Eq.(\ref{SUSYCS}).)
% on the bosonic body  manifolds (This is consistent with the 
%result in Subsect.\ref{SUSYHopfterm}).
However,  our SUSY CS theory still inherits topological natures peculiar to the 
 original CS theory and manifests them in a SUSY sense  as we shall see  in  Subsections \ref{Couplingwithmatter} and \ref{SUSYMCSandTPmass}.
Some comments are added further.
One may find ``various kinds'' of SUSY CS theories in literature.
For instance, in Ref.\cite{hep-th/9612031}, the SUSY CS theory is referred to 
as the CS theory with 
 $OSp(1|2)$
$\it{gauge}$ symmetry.
In Ref.\cite{PhysRevD45}, the SUSY CS theory is referred to as the CS theory
 coupled to 
 super matter currents.
 In the above, we  derived a new SUSY CS Lagrangian, which possesses the
 $OSp(1|2)$  $\it{global}$ 
supersymmetry. 
The matrix analogue of our SUSY CS Lagrangian is found in 
 Ref.\cite{hep-th/0311005}, where the SUSY CS term 
 plays a crucial role for the realization of fuzzy superspheres in the supermatrix model.
 In the following subsections, we investigate   field theoretical aspects  of our SUSY CS theory.

%%%%%%%%%%%%%%%%%%%%%%%%%%%%%%%%%%%%%%%%%%%%%%%%%%%%%%%%%%%%%%%%%%%%%%%%%%%%%%
\subsection{$U(1)$ gauge symmetry}
%%%%%%%%%%%%%%%%%%%%%%%%%%%%%%%%%%%%%%%%%%%%%%%%%%%%%%%%%%%%%%%%%%%%%%%%%%%%%%%

The original CS Lagrangian (\ref{originalCS}) is  invariant under the $U(1)$ gauge 
transformation up to a total derivative term.
Similarly,  under  the $U(1)$ gauge transformation with the gauge function $\Lambda$
%%%%%%%%%%%%%%%%%%%%%%%%%%%%%%%%%%%%%%%%%%%%%%%%%%%%%%%%%%
\begin{equation}
(A_a,A_{\alpha})\rightarrow (A_a,A_{\alpha})
+(\partial_{a}\Lambda,\partial_{\alpha}\Lambda),
\label{AU1}
\end{equation}
%%%%%%%%%%%%%%%%%%%%%%%%%%%%%%%%%%%%%%%%%%%%%%%%%%%%%%%%
 $\mathcal{L}_{sCS}$  only yields the total derivative terms
%%%%%%%%%%%%%%%%%%%%%%%%%%%%%%%%%%%%%%%%%%%%%%%%%%%%%%%%%%
\begin{align} 
&\delta\mathcal{L}_{sCS}=\frac{\kappa}{2}(\partial_a(\Lambda F_a)+\partial_{\alpha}(\Lambda F_{\alpha}))\nonumber\\
&~~~~~~~~~= \frac{\kappa}{4}(\partial_a(\Lambda \epsilon_{abc} F_{bc})- 
  i\partial_{\alpha}(\Lambda (C\sigma_a)_{\alpha\beta} F_{a\beta})\nonumber\\
&~~~~~~~~~~~~+\frac{i}{2} \partial_a (\Lambda  (C\sigma_a)_{\alpha\beta} F_{\alpha\beta})),
\end{align}
%%%%%%%%%%%%%%%%%%%%%%%%%%%%%%%%%%%%%%%%%%%%%%%%%%%%%%%%%%%%
where the scalar Bianchi identity (\ref{superBianchiire}) was used.
Thus, $\mathcal{L}_{sCS}$ is invariant under the $U(1)$ 
 gauge transformation when the boundary contributions are  neglected.

\subsection{Coupling with matter}\label{Couplingwithmatter}

Next, we investigate the SUSY CS theory with 
matter fields, by adding 
the interaction term
%%%%%%%%%%%%%%%%%%%%%%%%%%%%%%%%%%%%%%%%%%%%%%%%%%%%%%%%%%
\begin{equation}
\mathcal{L}_I=A_a J_a +A_{\alpha}J_{\alpha}
\label{superinteraction}
\end{equation}
%%%%%%%%%%%%%%%%%%%%%%%%%%%%%%%%%%%%%%%%%%%%%%%%%%%%%%%%%%%%
to  $\mathcal{L}_{sCS}$, where  
 $(J_a,J_{\alpha})$ are the matter currents that form a super multiplet     
%%%%%%%%%%%%%%%%%%%%%%%%%%%%%%%%%%%%%%%%%%%%%%%%%%%%%%%%%%%%%
\begin{subequations}
\begin{align}
&\delta_{\xi}J_a=\frac{1}{2}J_{\alpha}(\sigma_a \xi)_{\alpha},\\
&\delta_{\xi}J_{\alpha}=\frac{1}{2}J_a(C\sigma_a\xi)_{\alpha}.
\end{align}\label{superinteqmo}
\end{subequations}
%%%%%%%%%%%%%%%%%%%%%%%%%%%%%%%%%%%%%%%%%%%%%%%%%%%%%%%%%%%%%
The equations of motion  for SUSY CS fields are given by 
%%%%%%%%%%%%%%%%%%%%%%%%%%%%%%%%%%%%%%%%%%%%%%%%%%%%%%%%%%%
\begin{subequations}
\begin{align}
&{\kappa}F_a=-J_a,\\
&{\kappa}F_{\alpha}=-J_{\alpha}.
\end{align}\label{EQMforscurrentCS}
\end{subequations}
%%%%%%%%%%%%%%%%%%%%%%%%%%%%%%%%%%%%%%%%%%%%%%%%%%%%%%%%%%%
It is obvious that  the current conservation 
%%%%%%%%%%%%%%%%%%%%%%%%%%%%%%%%%%%%%%%%%%%%%%%%%%%%%%%%%%%%
\begin{equation}
\partial_a J_a +\partial_{\alpha} J_{\alpha}=0
\label{supcurrentcons}
\end{equation}
%%%%%%%%%%%%%%%%%%%%%%%%%%%%%%%%%%%%%%%%%%%%%%%%%%%%%%%%%%
 is compatible with the scalar super Bianchi identity (\ref{superBianchiire}).
Regarding the 3rd bosonic axis as the temporal direction, we consider 
  point particles coupled to the 
 SUSY CS fields in Wick-rotated Euclidean super space-time $\mathbb{R}^{3|2}$.
The Lagrangian is given by    
%%%%%%%%%%%%%%%%%%%%%%%%%%%%%%%%%%%%%%%%%%%%%%%%%%%%%%%%%%%%%%%%
\begin{equation}
L=\sum_p \frac{m}{2}( {\dot{x}_i^p}{\dot{x}_i^p}
+C_{\alpha\beta}\dot{\theta}_{\alpha}^p\dot{\theta}_{\beta}^p)+
\int \!\! d^2 x d^2\theta~ 
\mathcal{L}_I + \int\!\! d^2 x d^2\theta ~\mathcal{L}_{sCS},
\end{equation}
%%%%%%%%%%%%%%%%%%%%%%%%%%%%%%%%%%%%%%%%%%%%%%%%%%%%%%%%%%%%%%%%
where $(x_i^p,\theta_{\alpha}^p)$ $(i=1,2)$ denotes the position of the $p$-th particle, and 
 the matter currents are   
%%%%%%%%%%%%%%%%%%%%%%%%%%%%%%%%%%%%%%%%%%%%%%%%%%%%%%%%%%%%%%%%%%%%%%%%%%
\begin{subequations}
\begin{align}
&J_a (x,\theta)=\sum_p \dot{x}_a^p\delta^2(x-x^p)\delta^2(\theta-\theta^p),\\
&J_{\alpha}(x,\theta)=\sum_p \dot{\theta}_a^p\delta^2(x-x^p)\delta^2(\theta-\theta^p).
\end{align}\label{supermatterparticle}
\end{subequations} 
%%%%%%%%%%%%%%%%%%%%%%%%%%%%%%%%%%%%%%%%%%%%%%%%%%%%%%%%%%%%%%%%%%%%%%%%%%%%%
The equations of motion for the $p$-th particle are derived  as 
%%%%%%%%%%%%%%%%%%%%%%%%%%%%%%%%%%%%%%%%%%%%%%%%%%%%%%%%%%%%%%%%%%%%%%%%%
\begin{subequations}
\begin{align}
& m\ddot{x}_i^p=F_{ia}\dot{x}_a^p-F_{i\alpha}\dot{\theta}_{\alpha}^p,\\
& m\ddot{\theta}_{\alpha}^p=C_{\alpha\beta}(F_{a\beta}\dot{x}_a^p+F_{\beta\gamma}\dot{\theta}_{\gamma}^p).
\end{align}\label{ptheparticleEOM}
\end{subequations}
%%%%%%%%%%%%%%%%%%%%%%%%%%%%%%%%%%%%%%%%%%%%%%%%%%%%%%%%%%%%%%%%%%%%%%%
With the use of Eqs.(\ref{EQMforscurrentCS}), 
 the equations of motion (\ref{ptheparticleEOM}) are reduced to those of  the free particle 
%%%%%%%%%%%%%%%%%%%%%%%%%%%%%%%%%%%%%%%%%%%%%%%%%%%%%%%%%%%%%%%%%%%%
\begin{subequations}
\begin{align}
&\ddot{x}_i^p=0,\\
&\ddot{\theta}^p_{\alpha}=0.
\end{align}
\end{subequations}
%%%%%%%%%%%%%%%%%%%%%%%%%%%%%%%%%%%%%%%%%%%%%%%%%%%%%%%%%%%%%%% 
Thus, the SUSY CS gauge fields never affect the 
dynamical motion of the matter particles in  bulk.
This is a consequence of the topological nature of the SUSY CS theory.

\subsection{SUSY Hopf term}\label{SUSYHopfterm}

The phase interaction between 
 two anyons is described by the non-local 
Hopf term 
%%%%%%%%%%%%%%%%%%%%%%%%%%%%%%%%%%%%%%%%%%%%%%%%%%%%%%%%%%%%%%%%%%%%%%%%%
\begin{align}
&\mathcal{L}_{Hopf}=-\frac{1}{2\kappa}\int d^3 x \int d^3 y  ~\epsilon_{abc} J_a(x) 
\frac{x_b-y_b}{|x-y|^3}
   J_c(y)\nonumber\\
&~~~~~~~~= -\frac{1}{2\kappa}J\frac{\epsilon\partial}{\partial^2}J,
\label{orihopf}
\end{align}
%%%%%%%%%%%%%%%%%%%%%%%%%%%%%%%%%%%%%%%%%%%%%%%%%%%%%%%%%%%%%%%%%%%%%%%%% 
where $(\epsilon\partial)_{ab}=\epsilon_{abc}\partial_c$, $\partial^2=\partial_a^2$ and 
 $J_a$ represents the anyon current. 
Introducing the CS gauge fields,  
the  Hopf interaction is rewritten as  the
  CS local interaction 
$\frac{\kappa}{2} A \epsilon\partial A+A J.$
(Integrating out the CS fields, we obtain the Hopf term (\ref{orihopf}).)
The Hopf term corresponds to the topological invariant of the two closed (anyon) loops, known as 
Gauss linking number.
It would be quite interesting to see 
 the SUSY extension of 
the   Hopf term by integrating out the SUSY CS gauge 
fields in the SUSY CS Lagrangian.
For  later convenience, we rewrite the
 SUSY CS Lagrangian in  
 matrix representation  
%%%%%%%%%%%%%%%%%%%%%%%%%%%%%%%%%%%%%%%%%%%%%%%%%%%%%%%%%%%%%%%%%%%
\begin{align}
&\mathcal{L}=\mathcal{L}_{sCS}+\mathcal{L}_{I}\nonumber\\
&~~=\frac{\kappa}{2}
\begin{pmatrix}
& \!\! A \\
& \!\! \mathcal{A}
\end{pmatrix}^t
X
\begin{pmatrix}
& \!\! A \\
& \!\! \mathcal{A}
\end{pmatrix}
+
\begin{pmatrix}
& \!\! A \\
& \!\! \mathcal{A}
\end{pmatrix}^t
\begin{pmatrix}
& \!\! J \\
& \!\! \mathcal{J}
\end{pmatrix}
,\label{sCSplusmatter}
\end{align}
%%%%%%%%%%%%%%%%%%%%%%%%%%%%%%%%%%%%%%%%%%%%%%%%%%%%%%%%%%%%%%%%%%%
where $A_a$ and $J_a$ represent the bosonic gauge fields and the bosonic currents, respectively,
 while $\mathcal{A}_{\alpha}$ and $\mathcal{J}_{\alpha}$
 represent the fermionic ones. The matrix $X$ is given by   
%%%%%%%%%%%%%%%%%%%%%%%%%%%%%%%%%%%%%%%%%%%%%%%%%%%%%%%%%%%%%%%%%%%
\begin{align}
&X= \begin{pmatrix}
M & P\\
-P^t & N
\end{pmatrix}
\nonumber\\
&~~~\equiv
\begin{pmatrix}
 -\epsilon_{abc}\partial_{c} & \frac{i}{2}(C\sigma_a)_{\alpha\beta}\partial_{\alpha}\\
 \frac{i}{2}(C\sigma_b)_{\alpha\beta}\partial_{\beta} & -\frac{i}{2}(C\sigma_a)_{\alpha\beta}\partial_a\\
\end{pmatrix}.
\end{align}
%%%%%%%%%%%%%%%%%%%%%%%%%%%%%%%%%%%%%%%%%%%%%%%%%%%%%%%%%%%%%%%%%%%
Because of  the  existence of the $U(1)$ gauge freedom
(\ref{AU1}), with any smooth function $\Lambda$, the zero-mode equation of $X$ holds
%%%%%%%%%%%%%%%%%%%%%%%%%%%%%%%%%%%%%%%%%%%%%%%%%%%%%%%%%%%%%%%%%%%%%%
\begin{equation}
X\cdot 
\begin{pmatrix}
& \!\!\partial_a \Lambda\\
& \!\!\partial_{\alpha} \Lambda
\end{pmatrix}
=0.
\end{equation}
%%%%%%%%%%%%%%%%%%%%%%%%%%%%%%%%%%%%%%%%%%%%%%%%%%%%%%%%%%%%%%%%%%%%%%%%%
Then,  when we take the inverse of $X$,   the $U(1)$ gauge  freedom needs to be fixed
 by restricting the function space  
 in which $X$ does not have its zero-mode.
For instance, we may apply the super Lorentz gauge %%%%%%%%%%%%%%%%%%%%%%%%%%%%%%%%%%%%%%%%%%%%%%%%%%%%%%%%%%%%%%%%%%%%%%%%%%%  
\begin{equation}
\partial_a A_a +C_{\alpha\beta} \partial_{\alpha} A_{\beta}=0,
\end{equation}
%%%%%%%%%%%%%%%%%%%%%%%%%%%%%%%%%%%%%%%%%%%%%%%%%%%%%%%%%%%%%%%%%%%%%%%%
or the ordinary Lorentz gauge 
%%%%%%%%%%%%%%%%%%%%%%%%%%%%%%%%%%%%%%%%%%%%%%%%%%%%%%%%%%%%%%%%%%%%%%%%
\begin{equation}
\partial_a A_a=0.
\label{orilorgauge}
\end{equation}
%%%%%%%%%%%%%%%%%%%%%%%%%%%%%%%%%%%%%%%%%%%%%%%%%%%%%%%%%%%%%%%%%%
Integrating out the super gauge fields $(A,\mathcal{A})$ in Eq.(\ref{sCSplusmatter}), 
 a  SUSY extension of the  Hopf term is obtained as 
%%%%%%%%%%%%%%%%%%%%%%%%%%%%%%%%%%%%%%%%%%%%%%%%%%%%%%%%%%%%%%%
\begin{align}
&\mathcal{L}_{sHopf}=-\frac{1}{2\kappa}
\begin{pmatrix}
J\\
\mathcal{J}
\end{pmatrix}^t
\frac{1}{X}
\begin{pmatrix}
J\\
-\mathcal{J}
\end{pmatrix}
\nonumber\\
&~~~~~~~~=
-\frac{1}{2\kappa}
 \begin{pmatrix}
 J\\
\mathcal{J}
\end{pmatrix}^t
\begin{pmatrix}
 \tilde{M} & \tilde{M}P\frac{1}{N} \\
\frac{1}{N}P^t\tilde{M} & -\tilde{N}
\end{pmatrix}
\begin{pmatrix}
 J\\
\mathcal{J}
\end{pmatrix}\nonumber\\
&~~~~~~~~ =-\frac{1}{2\kappa}
 \begin{pmatrix}
 J\\
\mathcal{J}
\end{pmatrix}^t
\begin{pmatrix}
\tilde{M} & \frac{1}{M}P\tilde{N} \\
\tilde{N}P^t\frac{1}{M} & - \tilde{N}
\end{pmatrix}
\begin{pmatrix}
 J\\
\mathcal{J}
\end{pmatrix},
\label{superhopfinverse}
\end{align}
%%%%%%%%%%%%%%%%%%%%%%%%%%%%%%%%%%%%%%%%%%%%%%%%%%%%%%%%%%%%%%% 
where  $\tilde{M},\tilde{N}$ are 
 given by 
%%%%%%%%%%%%%%%%%%%%%%%%%%%%%%%%%%%%%%%%%%%%%%%%%%%%%%%%%%%%%%%%%%
\begin{subequations}
\begin{align}
&\tilde{M}=\frac{1}{M+P\frac{1}{N}P^t},%=\frac{1}{M} -\frac{1}{M}P\frac{1}{N}P^t\frac{1}{M}
\\
&\tilde{N}= \frac{1}{N+P^t\frac{1}{M}P}.%=\frac{1}{N}   -\frac{1}{N}P^t\frac{1}{M}P\frac{1}{N}
\end{align}
\label{tildemn}
\end{subequations}
%%%%%%%%%%%%%%%%%%%%%%%%%%%%%%%%%%%%%%%%%%%%%%%%%%%%%%%%%%%%%%%%%%
See also the formulas (\ref{inversexformu}) and (\ref{xinverse}) 
about the inverse of  the supermatrix. 
 Expanding $\tilde{M}$ and  $\tilde{N}$ in terms of 
 Grassmann odd quantity $P$, 
  $\mathcal{L}_{sHopf}$ is simply rewritten as 
%%%%%%%%%%%%%%%%%%%%%%%%%%%%%%%%%%%%%%%%%%%%%%%%%%%%%%%%%%%%%%%%%
\begin{align}
&\mathcal{L}_{sHopf}\nonumber\\
&=-\frac{1}{2\kappa}
\begin{pmatrix}
J\\
\mathcal{J}
\end{pmatrix}^t
\begin{pmatrix}\frac{1}{M} -\frac{1}{M}P\frac{1}{N}P^t\frac{1}{M} &  \frac{1}{M}P\frac{1}{N}  \\ 
 \frac{1}{N}P^t\frac{1}{M}    & -\frac{1}{N}   +\frac{1}{N}P^t\frac{1}{M}P\frac{1}{N}      
\end{pmatrix}
\begin{pmatrix}
J\\
\mathcal{J}
\end{pmatrix}\!\!.
\end{align}
%%%%%%%%%%%%%%%%%%%%%%%%%%%%%%%%%%%%%%%%%%%%%%%%%%%%%%%%%%%%%%%%%%%%%%%
Because of  the 
nilpotency of the Grassmann number,  terms more than quadratic about $P$ 
 do not appear in the expansion. 
In  the original derivative expression, block components of $X^{-1}$ are represented as   
%%%%%%%%%%%%%%%%%%%%%%%%%%%%%%%%%%%%%%%%%%%%%%%%%%%%%%%%%%%%%%%%%%%%%%
\begin{subequations}
\begin{align}
&\biggl(\frac{1}{M}-\frac{1}{M}P\frac{1}{N} {P}^t\frac{1}{M}\biggr)_{ab}\nonumber\\
&~~~~
= 
\biggl(\frac{\epsilon\partial}{\partial^2}\biggr)_{ab} -\frac{1}{2}
\biggl(\frac{\epsilon\partial}{\partial^2}\biggr)_{ab}^3 
  \frac{\partial}{\partial\theta}C \frac{\partial}{\partial\theta}, \label{lhopf11comp}
\\
& \biggl(-\frac{1}{N}+\frac{1}{N}P^t\frac{1}{M}P\frac{1}{N}\biggr)_{\alpha\beta}
\nonumber\\
&~~~~=
2i(\sigma_a C)_{\alpha\beta}\frac{\partial_a}{\partial^2}-i\biggl(\biggl(\frac
{\sigma_a\partial_a}{\partial^2}\biggr)^3 C\biggr)_{\alpha\beta}
\frac{\partial}{\partial\theta}C \frac{\partial}{\partial\theta}
,\\
&\biggl(\frac{1}{M}P\frac{1}{N}\biggr)_{a\alpha}
= 
-\epsilon_{abc}(\sigma_d\sigma_b)_{\alpha\beta}
\frac{\partial_c}{\partial^2}\frac{\partial_d}{\partial^2}\partial_{\beta},
\\
&\biggl(\frac{1}{N}P^t\frac{1}{M}\biggr)_{\alpha a}
=
-\epsilon_{abc}
(\sigma_d\sigma_b)_{\alpha\beta} \frac{\partial_d}{\partial^2}
 \frac{\partial_c}{\partial^2} \partial_{\beta},
\end{align} \label{hopfeachterm}
\end{subequations}
%%%%%%%%%%%%%%%%%%%%%%%%%%%%%%%%%%%%%%%%%%%%%%%%%%%%%%%%%%%%%%%%
where 
  $\frac{1}{\partial_a}=\frac{\partial_a}{\partial^2}$, 
$\partial^2=\partial_a^2$, $\delta^2(\theta)=\frac{1}{2}\theta C\theta$, $(\epsilon\partial)_{ab}
=\epsilon_{abc}\partial_c$  
and  we used the Lorentz gauge (\ref{orilorgauge}) to obtain   $\frac{1}{\epsilon\partial}=
-\frac{\epsilon\partial}{\partial^2}$. 
The identity matrix  $1$ corresponds to $\delta^3(x)\delta^2(\theta)$ in the function sense.
Since each inverse matrix performs one integration  in the parameter space,
  the SUSY Hopf term is  expressed by highly non-local interactions. 
For instance, the second term on  the right-hand side in Eq.(\ref{lhopf11comp}) contains 
$(\epsilon\partial/\partial^2)^3$, 
which performs $x$-space integration 3 times.
The fermionic terms in $\mathcal{L}_{sHopf}$ 
are regarded as  newly induced 
  phases by the 
 fermionic gauge field $\mathcal{A}$.
As the topological charge of the supermonopole is given by that of its bosonic counterpart 
\cite{math-ph/9907020}, we may define the knot invariant of the SUSY Hopf term by
 its original Hopf term which appears as the first term in Eq.(\ref{lhopf11comp}).

\subsection{ SUSY Maxwell-Chern-Simons theory and topological masses}\label{SUSYMCSandTPmass}

 The inner product of the 
  super vector field strengths  $(F_a, F_{\alpha})$ gives an $OSp(1|2)$
 singlet 
%%%%%%%%%%%%%%%%%%%%%%%%%%%%%%%%%%%%%%%%%%%%%%%%%%%%%%%%%%%%%%%%
\begin{align}
&\mathcal{L}_{sM}= -\frac{1}{2e^2}(F_a^2+C_{\alpha\beta}F_{\alpha}F_{\beta}) \nonumber\\
&~~~~~~= -\frac{1}{4e^2}(F_{ab}^2+\frac{i}{2}\epsilon_{abc}(C\sigma_c)(F_{ab}F_{\alpha\beta}+F_{a\alpha}F_{b\beta})\nonumber\\
&~~~~~~~~~~+\frac{1}{2}C_{\alpha\beta}F_{a\alpha}F_{a\beta}+\frac{1}{4}C_{\alpha\beta}C_{\gamma\delta}      F_{\alpha\gamma}F_{\beta\delta}),
\label{susyMax}
\end{align}
%%%%%%%%%%%%%%%%%%%%%%%%%%%%%%%%%%%%%%%%%%%%%%%%%%%%%%%%%%%%%%%%
which we adopt  as the SUSY Maxwell Lagrangian. 
The
 SUSY Maxwell-CS Lagrangian is constructed by  the coupling of  $\mathcal{L}_{sM}$ and  $\mathcal{L}_{sCS}$:
%%%%%%%%%%%%%%%%%%%%%%%%%%%%%%%%%%%%%%%%%%%%%%%%%%%%%%%%%%%%%%%%%%%%%%%%%%%
\begin{equation}
\mathcal{L}_{sMCS}=-\frac{1}{2e^2}(F_a^2+C_{\alpha\beta}F_{\alpha}F_{\beta}) 
+\frac{\kappa}{2}(A_a F_a +A_{\alpha}F_{\alpha}).
\label{susymaxcslag}
\end{equation}
%%%%%%%%%%%%%%%%%%%%%%%%%%%%%%%%%%%%%%%%%%%%%%%%%%%%%%%%%%%%%%%%%%%%%%%%%%%%
The equations of motion  are derived as 
%%%%%%%%%%%%%%%%%%%%%%%%%%%%%%%%%%%%%%%%%%%%%%%%%%%%%%%%%%%%%%%%%%
\begin{subequations}
\begin{align}
&\frac{1}{e^2}\partial_b F_{ba}
-i\frac{3}{8e^2}\epsilon_{abc}(C\sigma_c)_{\alpha\beta}\partial_b 
F_{\alpha\beta}
-\frac{1}{4e^2}C_{\alpha\beta}\partial_{\alpha}F_{a\beta}
\nonumber\\
&~~~~~=\kappa \epsilon_{abc}F_{bc}
+i\frac{\kappa}{2}(C\sigma_a)_{\alpha\beta}F_{\alpha\beta},\\
&i\frac{3}{8e^2}\epsilon_{abc}(C\sigma_c)_{\alpha\beta}
\partial_{\beta}F_{ab}
-\frac{1}{4e^2}C_{\alpha\beta}(\partial_a F_{a\beta}
+C_{\gamma\delta}\partial_{\gamma}F_{\beta\delta})\nonumber\\
&~~~~~=i\kappa (C\sigma_a)_{\alpha\beta}F_{a\beta}.
\end{align}\label{maxwellcseqns}
\end{subequations}
%%%%%%%%%%%%%%%%%%%%%%%%%%%%%%%%%%%%%%%%%%%%%%%%%%%%%%%%%%%%%%%%
With the use of  $(F_a, F_{\alpha})$, Eqs.(\ref{maxwellcseqns})
 are simply rewritten as 
%%%%%%%%%%%%%%%%%%%%%%%%%%%%%%%%%%%%%%%%%%%%%%%%%%%%%%%%%%%%%%%%
\begin{subequations}
\begin{align}
& \frac{1}{e^2}\epsilon_{abc}\partial_b F_c-\frac{i}{2e^2}(C\sigma_a C)_{\alpha\beta}\partial_{\alpha}F_{\beta}=-2\kappa F_a,\\
& \frac{i}{2e^2}(C\sigma_a C)_{\alpha\beta}\partial_a F_{\beta}+\frac{i}{2e^2}(C\sigma_a)_{\alpha\beta}\partial_{\beta}F_{a}=-2\kappa F_{\alpha}.
\end{align}\label{maxwellcseqns2}
\end{subequations}
%%%%%%%%%%%%%%%%%%%%%%%%%%%%%%%%%%%%%%%%%%%%%%%%%%%%%%%%%%%%%%%%%
 
It is well known that the CS term coupled to  the Maxwell Lagrangian induces a 
topological mass to gauge fields.
The mechanism is intuitively explained with the use of the 
analogy between the Maxwell-CS mechanics and the particle 
 mechanics  under  magnetic field \cite{PhysRevD41}. 
The CS term acts as ``Lorentz force'' in Maxwell-CS mechanics,
 and yields a cyclotron frequency which 
 corresponds to the gauge mass in  Maxwell-CS field theory.
Here, we perform  similar analyses in the SUSY Maxwell-CS  theory.
For this end, we treat the 3rd bosonic 
coordinate as the temporal direction, and work in the super space-time. 
In a low energy limit, the spatial derivatives of the gauge fields are dropped and the super 
vector field strengths are reduced to 
%%%%%%%%%%%%%%%%%%%%%%%%%%%%%%%%%%%%%%%%%%%%%%%%%%%%%%%%%%%%%%%%%%%%%%%%
\begin{equation}
F_i\rightarrow -\epsilon_{ij}\dot{A}_j, ~~~~ 
F_{\alpha}\rightarrow \frac{i}{2}(\sigma_1)_{\alpha\beta}\dot{A}_{\beta},
\end{equation}
%%%%%%%%%%%%%%%%%%%%%%%%%%%%%%%%%%%%%%%%%%%%%%%%%%%%%%%%%%%%%%%%%%%%%%%%
and the  Lagrangian for the SUSY Maxwell-CS mechanics is obtained as   
%%%%%%%%%%%%%%%%%%%%%%%%%%%%%%%%%%%%%%%%%%%%%%%%%%%%%%%%%%%
\begin{align}
&L_{sMCS}=\frac{1}{2e^2}\dot{A}^2_i+\frac{1}{8e^2}C_{\alpha\beta}\dot{A}_{\alpha}\dot{A}_{\beta}\nonumber\\
&~~~~~~~~~~~-\frac{\kappa}{2}\epsilon_{ij}A_i \dot{A}_j
+\frac{\kappa}{4}i(\sigma_1)_{\alpha\beta}A_{\alpha}\dot{A}_{\beta}.
\end{align}
%%%%%%%%%%%%%%%%%%%%%%%%%%%%%%%%%%%%%%%%%%%%%%%%%%%%%%%%%% 
The equations of motion (\ref{maxwellcseqns}) become
 %%%%%%%%%%%%%%%%%%%%%%%%%%%%%%%%%%%%%%%%%%%%%%%%%%%%%%%%%%%%%%%%%%
\begin{subequations}
\begin{align}
& \ddot{A}_i= e^2\kappa \epsilon_{ij}\dot{A}_j,\\   
&\ddot{A}_{\alpha}=-2ie^2\kappa (\sigma_3)_{\alpha\beta} \dot{A}_{\beta}.
\end{align}
\label{eqforsusymcsm}
\end{subequations}
%%%%%%%%%%%%%%%%%%%%%%%%%%%%%%%%%%%%%%%%%%%%%%%%%%%%%%%%%%%%%%%%%
$L_{sMCS}$ is formally equivalent to the 
 one-particle Lagrangian on the superplane (\ref{planarsusylag}) with  replacement 
%%%%%%%%%%%%%%%%%%%%%%%%%%%%%%%%%%%%%%%%%%%%%%%%%%%%%%%%%%%%%
\begin{align}
&(A_i,A_{\alpha})\rightarrow (x_i,2C_{\alpha\beta}\theta_{\beta}),\nonumber\\
&e^2\rightarrow 1/m,\nonumber\\
&\kappa \rightarrow -B.
\end{align}
%%%%%%%%%%%%%%%%%%%%%%%%%%%%%%%%%%%%%%%%%%%%%%%%%%%%%%%%%%%%%%%%%%
From Eqs.(\ref{eqforsusymcsm}),
 the cyclotron frequencies in SUSY Maxwell-CS mechanics 
 are given by   $\omega_B=e^2\kappa$ and  $\omega_F=2e^2 \kappa$, 
which, respectively, correspond to 
 the  topological masses for the  bosonic  and the fermionic gauge fields
 in the SUSY Maxwell-CS  field theory.

%%%%%%%%%%%%%%%%%%%%%%%%%%%%%%%%%%%%%%%%%%%%%%%%%%%%%%%%%%%%%
\section{The SUSY CSLG theory and its dual 
 representation}\label{dualTPsec}
%%%%%%%%%%%%%%%%%%%%%%%%%%%%%%%%%%%%%%%%%%%%%%%%%%%%%%%%%%%%%%%

In this section, we present an effective field theory  for   SUSY QHE.
We explore the charge-flux duality in  the SUSY QH system, and construct its dual description,
  where topological solitons arise as fundamental excitations.
  Based on the dual SUSY CSLG description,
 we discuss physical properties of 
topological solitons on the SUSY QH liquid.
Since we deal with the spinless  matter fields, the 
topological solitons are realized as  the SUSY extension of  vortices.

\subsection{Charge-flux duality}

For the existence of the charge-flux duality, 
 space(-time) dimension is crucial.
In general dimensions, the field strengths behave as 2-rank antisymmetric tensor, while 
in  3-dimensional  space, thanks to the existence of the  3-rank antisymmetric tensor,
 the 2-rank field strengths are mapped to the 3-vector field strengths  
%%%%%%%%%%%%%%%%%%%%%%%%%%%%%%%%%%%%%%%%%%%%%%%%%%%%%%%%%%%%%%%%%%%%%%%%%%%%%%%%%%%%%%%%%%%
\begin{equation}
F_a=\frac{1}{2}\epsilon_{abc} F_{bc}.
\end{equation}
%%%%%%%%%%%%%%%%%%%%%%%%%%%%%%%%%%%%%%%%%%%%%%%%%%%%%%%%%%%%%%%%%%%%%%%%%%%%%%%%%%%%%%%%%%%
Then, in 3-dimensional space, there exists one-to-one correspondence, called the charge-flux duality, between 
3-vector currents and  3-vector field strengths
%%%%%%%%%%%%%%%%%%%%%%%%%%%%%%%%%%%%%%%%%%%%%%%%%%%%%%%%%%%%%%%%%
\begin{equation}
J_{a}
\leftrightarrow 
{F}_a=\frac{1}{2}\epsilon_{abc}F_{bc},
\end{equation}
%%%%%%%%%%%%%%%%%%%%%%%%%%%%%%%%%%%%%%%%%%%%%%%%%%%%%%%%%%%%%%%%%%
or, in their components,
%%%%%%%%%%%%%%%%%%%%%%%%%%%%%%%%%%%%%%%%%%%%%%%%%%%%%%%%%%%%%%%%%%
\begin{equation}
(\rho,J_i)\leftrightarrow (B,E_i).
\end{equation}
%%%%%%%%%%%%%%%%%%%%%%%%%%%%%%%%%%%%%%%%%%%%%%%%%%%%%%%%%%%%%%%%%%
In the SUSY case, there  exists an analogous correspondence 
between super vector currents and  super vector field strengths:  
%%%%%%%%%%%%%%%%%%%%%%%%%%%%%%%%%%%%%%%%%%%%%%%%%%%%%%%%%%%%%%%%%%%%%%
\begin{subequations}
\begin{align}
&J_{a} \leftrightarrow  
{F}_{a}= \frac{1}{2}\epsilon_{abc}{F}_{bc}
+i\frac{1}{4}(C\sigma_a)_{\alpha\beta}{F}_{\alpha\beta},\\
&J_{\alpha} \leftrightarrow  
{F}_{a}= -\frac{i}{2}(C\sigma_a)_{\alpha\beta}{F}_{a\beta}.
\end{align}
\end{subequations}
%%%%%%%%%%%%%%%%%%%%%%%%%%%%%%%%%%%%%%%%%%%%%%%%%%%%%%%%%%%%%%%%%%%%%
Based on this one-to-one relation, we introduce the 
 dual CS field strengths $(\tilde{F}_a,\tilde{F}_{\alpha})$
%%%%%%%%%%%%%%%%%%%%%%%%%%%%%%%%%%%%%%%%%%%%%%%%%%%%%%%%
\begin{subequations}
\begin{align}
&\tilde{F}_a=\frac{1}{2}\epsilon_{abc}\tilde{F}_{bc}
+i\frac{1}{4}(C\sigma_a)_{\alpha\beta}\tilde{F}_{\alpha\beta},\\
& \tilde{F}_{\alpha}= -\frac{i}{2}(C\sigma_a)_{\alpha\beta}\tilde{F}_{a\beta},
\end{align}
\end{subequations}
%%%%%%%%%%%%%%%%%%%%%%%%%%%%%%%%%%%%%%%%%%%%%%%%%%%%%%%%%%%
to match the super matter currents 
%%%%%%%%%%%%%%%%%%%%%%%%%%%%%%%%%%%%%%%%%%%%%%%%%%%%%%%%%%%
\begin{subequations}
\begin{align}
&\tilde{F}_a=-J_a,\\
& \tilde{F}_{\alpha}=-J_{\alpha}.
\end{align}\label{jtildef}
\end{subequations}
%%%%%%%%%%%%%%%%%%%%%%%%%%%%%%%%%%%%%%%%%%%%%%%%%%%%%%%%%%%
Since $(\tilde{F}_a,\tilde{F}_{\alpha})$ satisfy the super Bianchi identity $\partial_{a}\tilde{F}_a+\partial_{\alpha}\tilde{F}_{\alpha}=0$, Eqs.(\ref{jtildef}) are  
  compatible
with the current conservation $\partial_a J_a+\partial_{\alpha}J_{\alpha}=0$. 
With the use of the original and the dual CS  fields,
 the SUSY CS Lagrangian is expressed as  
%%%%%%%%%%%%%%%%%%%%%%%%%%%%%%%%%%%%%%%%%%%%%%%%%%%%%%%%%%%%%%%%%%%
\begin{align}
&\mathcal{L}= \mathcal{L}_{sCS}+\mathcal{L}_I\nonumber\\
&~~~=\frac{\kappa}{2}(A_a F_a +A_{\alpha} F_{\alpha})-A_a \tilde{F}_a-A_{\alpha}\tilde{F}_{\alpha}\nonumber\\
&~~~=\frac{\kappa}{4}(\epsilon_{abc}A_aF_{bc}-i(C\sigma_a)_{\alpha\beta}A_{\alpha}F_{a\beta}
+\frac{i}{2}(C\sigma_a)_{\alpha\beta}A_aF_{\alpha\beta})\nonumber\\
&~~~-\frac{1}{2}\epsilon_{abc}A_a\tilde{F}_{bc}
+\frac{i}{2}(C\sigma_a)_{\alpha\beta}A_{\alpha}\tilde{F}_{a\beta}
-\frac{i}{4}(C\sigma_a)_{\alpha\beta}A_a\tilde{F}_{\alpha\beta}.
\end{align}
%%%%%%%%%%%%%%%%%%%%%%%%%%%%%%%%%%%%%%%%%%%%%%%%%%%%%%%%%%%%%%%%%%
Integrating out the original CS fields $(A_a,A_{\alpha})$,
 we obtain the dual CS Lagrangian
%%%%%%%%%%%%%%%%%%%%%%%%%%%%%%%%%%%%%%%%%%%%%%%%%%%%%%%%%%%%%%%%%%%
\begin{align}
&\tilde{\mathcal{L}}_{sCS}=-\frac{1}{2\kappa}(\tilde{A}_a \tilde{F}_a+\tilde{A}_{\alpha}\tilde{F}_{\alpha})\nonumber\\
&~~= -\frac{1}{4\kappa}(\epsilon_{abc}\tilde{A}_a\tilde{F}_{bc}-i(C\sigma_a)_{\alpha\beta}\tilde{A}_{\alpha}\tilde{F}_{a\beta}
+\frac{i}{2}(C\sigma_a)_{\alpha\beta}\tilde{A}_a\tilde{F}_{\alpha\beta}).
\end{align}
%%%%%%%%%%%%%%%%%%%%%%%%%%%%%%%%%%%%%%%%%%%%%%%%%%%%%%%%%%%%%%%%%%%%%%%
Since the CS coupling constant in  $\tilde{\mathcal{L}}_{sCS}$ is  
 opposite to that in  $\mathcal{L}_{sCS}$, 
the charge-flux duality is restated as  the 
 $s$-dual transformation of   CS coupling.

%%%%%%%%%%%%%%%%%%%%%%%%%%%%%%%%%%%%%%%%%%%%%%%%%%%%%%%%%%
\subsection{Relativistic SUSY CSLG theory}\label{supervrela}
%%%%%%%%%%%%%%%%%%%%%%%%%%%%%%%%%%%%%%%%%%%%%%%%%%%%%%%%%%

Before detail discussions of  the nonrelativistic CSLG theory for SUSY QH liquid, 
it would be worthwhile to explore  the relativistic 
 formulation of  SUSY CSLG theory.
We regard the 3rd bosonic axis as the temporal direction, and deal with a 
 covariant SUSY CSLG theory  in the Wick-rotated 
 super space-time $\mathbb{R}^{3|2}$.
The essential features of duality  can be found in this relativistic formulation.
We introduce an
 complex scalar field coupled to  the SUSY CS fields as
%%%%%%%%%%%%%%%%%%%%%%%%%%%%%%%%%%%%%%%%%%%%%%%%%%%%%%%%
\begin{align}
&\mathcal{L}_{CSLG}
= (\partial_a+ic_a) \phi^* \cdot (\partial_a-ic_a) \phi\nonumber\\
&~~~~~+    C_{\alpha\beta} (\partial_{\alpha}+ic_{\alpha})   {\phi}^* \cdot(\partial_{\beta}-ic_{\beta}) \phi
\nonumber\\
&~~~~~+\frac{\kappa}{8\pi}(\epsilon_{abc}c_a f_{bc}-
i(C\sigma_a)_{\alpha\beta}c_{\alpha}f_{a\beta}+\frac{i}{2}(C\sigma_a)_{\alpha\beta}c_af_{\alpha\beta}),
\label{originalCSrel}
\end{align}
%%%%%%%%%%%%%%%%%%%%%%%%%%%%%%%%%%%%%%%%%%%%%%%%%%%%%%%%
where we denote the SUSY CS fields as 
$(c_a,c_{\alpha})$, and 
  their field strengths as  $(f_{ab},f_{a\alpha},f_{\alpha\beta})$.
The field equation for $\phi$ is given by 
%%%%%%%%%%%%%%%%%%%%%%%%%%%%%%%%%%%%%%%%%%%%%%%%%%%%%%%%%%%%%
\begin{equation}
(\partial_a-ic_a)^2\phi+C_{\alpha\beta}(\partial_{\alpha}-ic_{\alpha})(\partial_{\beta}-ic_{\beta})
\phi=0.
\end{equation}
%%%%%%%%%%%%%%%%%%%%%%%%%%%%%%%%%%%%%%%%%%%%%%%%%%%%%%%%%%%%%%
We decompose the complex scalar field into  the density part $\rho$ and the phase part 
$\chi$ as 
%%%%%%%%%%%%%%%%%%%%%%%%%%%%%%%%%%%%%%%%%%%%%%%%%%%%%%%
\begin{equation}
\phi=\sqrt{\rho}\chi.
\end{equation}
%%%%%%%%%%%%%%%%%%%%%%%%%%%%%%%%%%%%%%%%%%%%%%%%%%%%%%%%%
With the use of this decomposition,
 the kinetic term for $\phi$ is rewritten as  
%%%%%%%%%%%%%%%%%%%%%%%%%%%%%%%%%%%%%%%%%%%%%%%%%%%%%%%%%
\begin{align}
&\mathcal{L}_K=\rho[(i\chi^*(\partial_a-ic_a)\chi)^2
\nonumber\\
&~~~~~+C_{\alpha\beta}(i\chi^*(\partial_{\alpha}+ic_{\alpha})\chi\cdot i\chi^*
(\partial_{\beta}-ic_{\beta})\chi)],
\end{align}
%%%%%%%%%%%%%%%%%%%%%%%%%%%%%%%%%%%%%%%%%%%%%%%%%%%%%%%%%
where we postulated the density fluctuations  are very small.
With the use of the Stratonovich-Hubbard transformation formula \footnote{ 
Introducing the auxiliary fields
 $(J_a,J_{\alpha})$, the SUSY quadratic term $ V_a^2+C_{\alpha\beta}V_{\alpha}V_{\beta}$ is rewritten as 
$V_a J_a +V_{\alpha}J_{\alpha}-\frac{1}{4}
(J_a^2+C_{\alpha\beta}J_{\alpha}J_{\beta}).$},
%%%%%%%%%%%%%%%%%%%%%%%%%%%%%%%%%%%%%%%%%%%%%%%%%%%%%%%%%%%%%%%%%%%%%%%%
$\mathcal{L}_{K}$  is expressed as
%%%%%%%%%%%%%%%%%%%%%%%%%%%%%%%%%%%%%%%%%%%%%%%%%%%%%%%%%%%%%%
\begin{align}
&\mathcal{L}_K=(i\chi^*(\partial_a-ic_a)\chi)\cdot J_a+
(i\chi^*(\partial_{\alpha}-ic _{\alpha})\chi)\cdot J_{\alpha}\nonumber\\
&~~~~-\frac{1}{4\rho}(J_a^2+C_{\alpha\beta}J_{\alpha}J_{\beta}),
\label{auxikinetic}
\end{align}
%%%%%%%%%%%%%%%%%%%%%%%%%%%%%%%%%%%%%%%%%%%%%%%%%%%%%%%%%%%%%%%
where $(J_a,J_{\alpha})$ are auxiliary fields.
If there were not CS fields, the equations of motion for $(J_a,J_{\alpha})$
 would be given by  
%%%%%%%%%%%%%%%%%%%%%%%%%%%%%%%%%%%%%%%%%%%%%%%%%%%%%%%%%%%%%%%%%%%%%%%%%%%
\begin{subequations}
\begin{align}
&J_a=i\phi^* \overleftrightarrow{\partial_a} \phi,\\
&J_{\alpha}=iC_{\alpha\beta}\phi^*\overleftrightarrow{\partial_\beta} \phi.
\end{align}
\end{subequations}
%%%%%%%%%%%%%%%%%%%%%%%%%%%%%%%%%%%%%%%%%%%%%%%%%%%%%%%%%%%%%%%%%%%%%%%%%%
Thus, $(J_a,J_{\alpha})$  are 
  essentially the $U(1)$ conserved currents.
From Eq.(\ref{originalCSrel}), the equations of motion for 
 $(c_a, c_{\alpha})$ are given by 
%%%%%%%%%%%%%%%%%%%%%%%%%%%%%%%%%%%%%%%%%%%%%%%%%%%%%%%%%%%%%
\begin{subequations}
\begin{align}
&J_a=\frac{\kappa}{2\pi}f_a,
%~~~
\\
&J_{\alpha}=\frac{\kappa}{2\pi}f_{\alpha}.
\end{align}\label{fluxattchern}
\end{subequations}
%%%%%%%%%%%%%%%%%%%%%%%%%%%%%%%%%%%%%%%%%%%%%%%%%%%%%%%%%%

We further decompose the phase part into the smooth part and the singular part 
(vortex part) $\varphi$ as 
%%%%%%%%%%%%%%%%%%%%%%%%%%%%%%%%%%%%%%%%%%%%%%%%%%%%%%%%%%%%%
\begin{equation}
\chi=e^{-i\theta}\varphi,
\end{equation}
%%%%%%%%%%%%%%%%%%%%%%%%%%%%%%%%%%%%%%%%%%%%%%%%%%%%%
and   obtain 
%%%%%%%%%%%%%%%%%%%%%%%%%%%%%%%%%%%%%%%%%%%%%%%%%%%%%%%%%%%%
\begin{align}
&\mathcal{L}_K=(\partial_a\theta+  i\varphi^{*} {\partial_a}\varphi +c_a)J_a
+(\partial_{\alpha}\theta+ i\varphi^{*} {\partial_{\alpha}}\varphi  +c_{\alpha})J_{\alpha}\nonumber\\
&~~~~-\frac{1}{4\rho}(J_a^2+C_{\alpha\beta}J_{\alpha}J_{\beta}).
\end{align}
%%%%%%%%%%%%%%%%%%%%%%%%%%%%%%%%%%%%%%%%%%%%%%%%%%%
Integrating out the smooth function $\theta$, we find the 
 current conservation law 
$\partial_a J_a+\partial_{\alpha}J_{\alpha}=0$.
We introduce the dual CS fields $(\tilde{c}_a,\tilde{c}_{\alpha})$ 
based on the relation (\ref{jtildef}), and 
rewrite  the total Lagrangian $\mathcal{L}_{CSLG}$ as 
%%%%%%%%%%%%%%%%%%%%%%%%%%%%%%%%%%%%%%%%%%%%%%%%%%%%%%%%%%%%%
\begin{align}
&\mathcal{L}_{CSLG}= 
2\pi\tilde{c}_a\tilde{J}_a+2\pi \tilde{c}_{\alpha}\tilde{J}_{\alpha}
-c_a\tilde{f}_a-c_{\alpha}\tilde{f}_{\alpha}\nonumber\\
&~~~~~~~~~+\frac{\kappa}{8\pi}(c_a f_a+c_{\alpha}f_{\alpha})-\frac{1}{16\rho}(\tilde{f}_a^2+C_{\alpha\beta}\tilde{f}_{\alpha}\tilde{f}_{\beta}),
\label{CSdualCSLG}
\end{align}
%%%%%%%%%%%%%%%%%%%%%%%%%%%%%%%%%%%%%%%%%%%%%%%%%%%%%%%%%%%%%%
where we used a partial integration, and 
$(\tilde{J_a},\tilde{J}_{\alpha})$ denote the topological currents for vortex
%%%%%%%%%%%%%%%%%%%%%%%%%%%%%%%%%%%%%%%%%%%%%%%%%%%%%%%
\begin{subequations}
\begin{align}
&\tilde{J}_a= -\frac{i}{2\pi}\epsilon_{abc}\partial_b(\varphi^*\partial_c\varphi )
+\frac{1}{4\pi}(C\sigma_a)_{\alpha\beta}\partial_{\alpha}(\varphi^* \partial_{\beta}\varphi),\\
&\tilde{J}_{\alpha}=-\frac{1}{4\pi}(C\sigma_a)_{\alpha\beta}[\partial_a(\varphi^*\partial_{\beta}\varphi)-\partial_{\beta}(\varphi^*\partial_a\varphi)]. 
\end{align}\label{topologicalcurrentvortex}
\end{subequations}
%%%%%%%%%%%%%%%%%%%%%%%%%%%%%%%%%%%%%%%%%%%%%%%%%%%%%%%%%%%%%
Since  $(\tilde{J}_a,\tilde{J}_{\alpha})$ are topological currents, they 
automatically satisfy the 
current conservation law, $\partial_a \tilde{J}_a+\partial_{\alpha}\tilde{J}_{\alpha}=0$.

In the original bosonic CSLG theory,
the explicit representation of the vortex part  $\varphi$,  is given by 
 $\varphi(x)=\exp(i\alpha(x))$, where $\alpha(x)=\sum_p\alpha(x-x^p)=\sum_p
 arctan \frac{x_2-x^p_2}{x_1-x_1^p}$,  and 
 yields the topological charge density
%%%%%%%%%%%%%%%%%%%%%%%%%%%%%%%%%%%%%%%%%%%%%%%%%%%%%%%%%%%%%%
\begin{align}
&\tilde{\rho}(x)=-\frac{i}{2\pi}\epsilon_{ij}\partial_i(\varphi^{*}\partial_j\varphi)
=-\frac{1}{2\pi}\epsilon_{ij}\partial_i\partial_j\alpha(x)\nonumber\\
&~~~~~~=-\sum_p\delta^2(x-x^p),
\end{align}
%%%%%%%%%%%%%%%%%%%%%%%%%%%%%%%%%%%%%%%%%%%%%%%%%%%%%%%%%%%%%%%
 where $x^p$ denotes the position of the $p$-th vortex.
See Ref.\cite{hep-th/9202012} for detail discussions.
Similarly, in the SUSY case, the vortex part  $\varphi$ may be  expressed as  
 $\varphi(x,\theta)=\exp(i\alpha(x,\theta))$ with $\alpha(x,\theta)=\sum_p \alpha(x-x^p) \delta^2 (\theta-\theta^p)$, and yields the super topological charge density 
%%%%%%%%%%%%%%%%%%%%%%%%%%%%%%%%%%%%%%%%%%%%%%%%%%%%%%%%%%%%%%
\begin{align}
&\tilde{\rho}(x)=-\frac{i}{2\pi}\epsilon_{ij}\partial_i(\varphi^{*}\partial_j\varphi)
+\frac{1}{4\pi}(\sigma_1)_{\alpha\beta}\partial_{\alpha}(\varphi^*\partial_{\beta}\varphi)
\nonumber\\
&~~~~~~=-\frac{1}{2\pi}\epsilon_{ij}\partial_i\partial_j\alpha(x,\theta)+\frac{1}{2\pi}(\sigma_1)_{\alpha\beta}\partial_{\alpha}\partial_{\beta}\alpha(x,\theta)\nonumber\\
&~~~~~~=-\frac{1}{2\pi}\epsilon_{ij}\partial_i\partial_j\alpha(x,\theta)\nonumber\\
&~~~~~~=-\sum_p\delta^2(x-x^p)\delta^2(\theta-\theta^p).
\end{align}
%%%%%%%%%%%%%%%%%%%%%%%%%%%%%%%%%%%%%%%%%%%%%%%%%%%%%%%%%%%%%%
Integrating out the original SUSY CS fields $(c_a,c_{\alpha})$ in Eq.(\ref{CSdualCSLG}), the dual SUSY CSLG 
Lagrangian is derived as 
%%%%%%%%%%%%%%%%%%%%%%%%%%%%%%%%%%%%%%%%%%%%%%%%%%%%%%
\begin{align}
&\tilde{\mathcal{L}}_{CSLG}\nonumber\\
&~~= 
2\pi\tilde{c}_a 
\tilde{J}_a + 2\pi\tilde{c}_{\alpha} \tilde{J}_{\alpha}-\frac{\pi}{2\kappa}(\tilde{c}_a\tilde{f}_{a}+\tilde{c}_{\alpha}\tilde{f}_{\alpha})\nonumber\\
&~~-\frac{1}{16\rho}(\tilde{f}_{a}^2+C_{\alpha\beta}\tilde{f}_{\alpha}\tilde{f}_{\beta})
\nonumber\\
&~~=
2\pi\tilde{c}_a 
\tilde{J}_a + 2\pi\tilde{c}_{\alpha} \tilde{J}_{\alpha} 
\nonumber\\
&~~ -\frac{\pi}{2\kappa}(\epsilon_{abc}\tilde{c}_a\tilde{f}_{bc}
+i\frac{1}{2}(C\sigma_a)_{\alpha\beta}\tilde{c}_a\tilde{f}_{\alpha\beta}
-i(C\sigma_a)_{\alpha\beta}\tilde{c}_{\alpha}\tilde{f}_{a\beta})\nonumber\\
&~~-\frac{1}{8\rho}(\tilde{f}_{ab}^2+\frac{i}{2}\epsilon_{abc}(C\sigma_c)_{\alpha\beta}(\tilde{f}_{ab}\tilde{f}_{\alpha\beta}+\tilde{f}_{a\alpha}\tilde{f}_{b\beta})\nonumber\\
&~~~~~~+
\frac{1}{2}C_{\alpha\beta}\tilde{f}_{a\alpha}\tilde{f}_{a\beta}+\frac{1}{4}C_{\alpha\beta}C_{\gamma\delta}\tilde{f}_{\alpha\gamma}\tilde{f}_{\beta\delta}).
\end{align}
%%%%%%%%%%%%%%%%%%%%%%%%%%%%%%%%%%%%%%%%%%%%%%%%%%%%%%%
Thus, $\tilde{\mathcal{L}}_{CSLG}$ is equivalent to the SUSY 
 Maxwell-CS Lagrangian coupled to  topological currents.
The equations of motion for $(\tilde{c}_a,\tilde{c}_{\alpha})$ are derived as   
%%%%%%%%%%%%%%%%%%%%%%%%%%%%%%%%%%%%%%%%%%%%%%%%%%%%%%%%%%%%%%%
\begin{subequations}
\begin{align}
&\frac{1}{2}\epsilon_{abc}\partial_b \tilde{f}_c
+\frac{i}{4}(C\sigma_a )_{\alpha\beta}\partial_{\alpha}\tilde{f}_{\beta}
-\frac{2\rho\pi}{\kappa} \tilde{f}_a=-4\rho\pi \tilde{J}_a,\\
&-\frac{i}{4}(\sigma_a )_{\alpha\beta}\partial_a \tilde{f}_{\beta}
+\frac{i}{4}(\sigma_a)_{\alpha\beta}\partial_{\beta}\tilde{f}_a-\frac{2\rho\pi}{\kappa}
 \tilde{f}_{\alpha}=-4\rho\pi\tilde{J}_{\alpha}.
\end{align}\label{eqmotdualcs}
\end{subequations}
%%%%%%%%%%%%%%%%%%%%%%%%%%%%%%%%%%%%%%%%%%%%%%%%%%%%%%%%%%%%%
In a low energy limit, the higher derivatives in  SUSY Maxwell term are
 neglected, and  the dual SUSY CSLG Lagrangian is approximated by  the 
 super vortex Lagrangian coupled to 
 the dual SUSY  CS fields. 
Then,  replacing   vortex with matter, and the  
 CS coupling constant with its inverse, the dual SUSY CSLG Lagrangian has  the   
 same form as the original SUSY CSLG Lagrangian.
Similarly,  in a low energy limit, 
Eqs.(\ref{eqmotdualcs}) are reduced to 
%%%%%%%%%%%%%%%%%%%%%%%%%%%%%%%%%%%%%%%%%%%%%%%%%%%%%%%%%%%%%%%%%%%%%%%%%%%%%%
\begin{subequations}
\begin{align}
&\tilde{f}_a=2\kappa \tilde{J}_a,\\
&\tilde{f}_{\alpha}=2\kappa \tilde{J}_{\alpha}.
\end{align}\label{fluxattcherndual}
\end{subequations}
%%%%%%%%%%%%%%%%%%%%%%%%%%%%%%%%%%%%%%%%%%%%%%%%%%%%%%%%%%%%%%%%%%%%%%%%%%%%%%
Comparing Eqs.(\ref{fluxattcherndual}) with  Eqs.(\ref{fluxattchern}),
 one may find that the relation between charge and flux 
in the dual SUSY CSLG theory is opposite to that in 
the original SUSY CSLG theory.
It  confirms the previous observation that 
the $s$-dual transformation of the CS coupling  corresponds to the
 charge-flux duality.
Such dual feature has already been reported in the study of    
 the original bosonic CSLG theory \cite{PRL8862,IJMP92B6}, and 
 our SUSY CSLG theory shares it. 
This may be considered as  a manifestation that our SUSY CSLG theory provides a natural SUSY framework for 
the  original CSLG theory.

%%%%%%%%%%%%%%%%%%%%%%%%%%%%%%%%%%%%%%%%%%%%%%%%%%%%%%%%%%%
\subsection{Nonrelativistic  CSLG theory for  SUSY QH liquid }
\label{supervnonrela}
%%%%%%%%%%%%%%%%%%%%%%%%%%%%%%%%%%%%%%%%%%%%%%%%%%%%%%%%%%%

Here, we work in the super space-time again,  
and  construct the nonrelativistic SUSY CSLG theory on the superplane. 
The magnetic field and the electric fields are 
defined by the super vector fields $(F_a,F_{\alpha})$ 
as $B=-F_t,~ E_i=F_i ~(i=1,2),~ E_{\alpha}=F_{\alpha}$.
Since the original QH liquid is described by  composite bosons coupled to CS fields, 
 it would be reasonable to adopt the following  nonrelativistic  CSLG Lagrangian 
as the effective field theoretical description for the SUSY QH liquid     
%%%%%%%%%%%%%%%%%%%%%%%%%%%%%%%%%%%%%%%%%%%%%%%%%%%%%%
\begin{align}
&\mathcal{L}^{nr}_{CSLG}=
\phi^{*}(i\partial_t+\delta A_t)\phi\nonumber\\
&~~~~-\frac{1}{2m}
(-i\partial_i+\delta A_i) \phi^{*}\cdot (i\partial_i+\delta A_i)\phi\nonumber\\
&~~~~ -\frac{1}{2m}C_{\alpha\beta}(-i\partial_{\alpha}+\delta A_{\alpha})\phi^{*}
\cdot (i\partial_\beta+\delta A_{\beta}) \phi\nonumber\\
&~~~~-\frac{1}{2} \int  \delta \rho \!\cdot\! V\!\cdot \!\delta \rho\nonumber\\
&~~~~+\frac{\nu}{8\pi}(\epsilon_{abc}c_a f_{bc}
-i(C\sigma_a)_{\alpha\beta} c_{\alpha}f_{a\beta}
+\frac{i}{2}(C\sigma_a)_{\alpha\beta}c_a f_{\alpha\beta} ),
\end{align}
%%%%%%%%%%%%%%%%%%%%%%%%%%%%%%%%%%%%%%%%%%%%%%%%%%%%%%%%%
where $\phi$ denotes the composite boson field, 
 $\delta \rho = \rho-\bar{\rho}  $ is the net charge on the background,  $\nu=2\pi\bar{\rho}/B$
 is the filling factor, and 
 $(\delta A_a,\delta A_{\alpha})$ are the effective gauge fields 
for  composite bosons
%%%%%%%%%%%%%%%%%%%%%%%%%%%%%%%%%%%%%%%%%%%%%%%%%%%%%%%%%%%%
\begin{subequations}
\begin{align}
&\delta A_a=A_a-c_a,\\
&\delta A_{\alpha}=A_{\alpha}-c_{\alpha},
\end{align}
\end{subequations}
%%%%%%%%%%%%%%%%%%%%%%%%%%%%%%%%%%%%%%%%%%%%%%%%%%%%%%%%%%%
 where $(A_a,A_{\alpha})$ are the external gauge fields.

Since  $\mathcal{L}_{CSLG}^{nr}$ does not possess a  quadratic term about  time derivative,
  the Stratonovich-Hubbard transformations are  applied only
  to  quadratic terms about 
the  (super)spatial derivatives.
Except for this step, following procedures similar to Sec.\ref{supervrela}, 
  the dual 
nonrelativistic SUSY CSLG Lagrangian is derived as  
%%%%%%%%%%%%%%%%%%%%%%%%%%%%%%%%%%%%%%%%%%%%%%%%%%%%%%%%%%%%%%%%%%
\begin{align}
&\tilde{\mathcal{L}}_{CSLG}^{nr}=2\pi(-\tilde{c}_t\tilde{J}_t+\tilde{c}_i \tilde{J}_i+\tilde{c}_{\alpha}\tilde{J}_{\alpha})\nonumber\\
&~~~~-\frac{\pi}{2\nu}(\epsilon^{abc}\tilde{c}_a(\tilde{f}_{bc}+ \frac{\nu}{\pi} F_{bc})
-i(C\sigma^a)_{\alpha\beta}\tilde{c}_{\alpha}(\tilde{f}_{a\beta}+ \frac{\nu}{\pi} F_{a\beta})\nonumber\\
&~~~~~~~~~+\frac{i}{2}(C\sigma^a)_{\alpha\beta}\tilde{c}_a(\tilde{f}_{\alpha\beta}+\frac{\nu}{\pi}  F_{\alpha\beta}))\nonumber\\
&~~~~-\frac{m}{2\rho}(\tilde{f}_{ti}^2+\frac{i}{2}\epsilon_{ij}(C\sigma_i)_{\alpha\beta}
\tilde{f}_{tj}\tilde{f}_{\alpha\beta}\nonumber\\
&~~~~~~~~~+\frac{1}{16}(C\sigma_i)_{\alpha\beta}(C\sigma_i)_{\gamma\delta}\tilde{f}_{\alpha\beta}\tilde{f}_{\gamma\delta}+\frac{1}{4}(C\sigma^a\sigma^b)_{\alpha\beta}\tilde{f}_{a\alpha}\tilde{f}_{b\beta})\nonumber\\
&~~~~-\frac{1}{2}\int \delta \rho \!\cdot\! V \!\cdot\! \delta \rho,
\label{dualLagrangevortex}
\end{align}
%%%%%%%%%%%%%%%%%%%%%%%%%%%%%%%%%%%%%%%%%%%%%%%%%%%%%%%%%%%%%%%%%%
where $\delta \rho$ is given by  $\delta\rho=\frac{1}{2}\epsilon_{ij}(\tilde{f}_{ij}+\frac{\nu}{2\pi}F_{ij} ) +\frac{i}{4}(\sigma_1)_{\alpha\beta}( \tilde{f}_{\alpha\beta}+\frac{\nu}{2\pi}F_{\alpha\beta})$.
Further, we integrate out $\tilde{c}_t$ to obtain an effective action for
  super vortices. 
The equation of motion for $\tilde{c}_t$ is given by
%%%%%%%%%%%%%%%%%%%%%%%%%%%%%%%%%%%%%%%%%%%%%%%%%%%%%%%%%%%%%%%%%%%
\begin{equation}
( \partial_i^2+\frac{1}{4}C_{\alpha\beta}\partial_{\alpha}\partial_{\beta})\tilde{c}_t
=-\frac{2\pi\rho}{m}(\tilde{\rho}-\frac{1}{\nu}\delta\rho)+\text{(higher derivatives)},
\end{equation}
%%%%%%%%%%%%%%%%%%%%%%%%%%%%%%%%%%%%%%%%%%%%%%%%%%%%%%%%%%%%%%%%%%%
where $\tilde{\rho}=-\tilde{J}_t$.
Eliminating $\tilde{c}_t$ in Eq.(\ref{dualLagrangevortex}), the density excess is given by 
%%%%%%%%%%%%%%%%%%%%%%%%%%%%%%%%%%%%%%%%%%%%%%%%%%%%%%%%%%%%%
\begin{equation}
\delta\rho=\nu\tilde{\rho}.
\label{tpelecharge}
\end{equation}
%%%%%%%%%%%%%%%%%%%%%%%%%%%%%%%%%%%%%%%%%%%%%%%%%%%%%%%%%%%%%
This relation suggests that the super vortex with unit topological number carries the fractional 
electric charge $\nu$. 
Inserting Eq.(\ref{tpelecharge}) to  $\tilde{\mathcal{L}}_{CSLG}^{nr}$ 
and extracting vortex part, 
 the effective Lagrangian for the super vortex is obtained as 
%%%%%%%%%%%%%%%%%%%%%%%%%%%%%%%%%%%%%%%%%%%%%%%%%%%%%%%%%%%%%
\begin{equation}
\tilde{\mathcal{L}}_{eff}=2\pi(\tilde{c}_i \tilde{J}_i+\tilde{c}_{\alpha}\tilde{J}_{\alpha})+\nu A_t\tilde{\rho} -\frac{1}{2}\nu^2\int  \tilde{\rho}\!\cdot\! V \!\cdot\! \tilde{\rho}.
\end{equation}
%%%%%%%%%%%%%%%%%%%%%%%%%%%%%%%%%%%%%%%%%%%%%%%%%%%%%%%%%%%%
As expected, the super vortex is coupled to the electric potential with coupling $\nu$.
In a low energy limit in which super vortices are approximated  as  point-like objects,
 $\tilde{\mathcal{L}}_{eff}$
is  written as  
%%%%%%%%%%%%%%%%%%%%%%%%%%%%%%%%%%%%%%%%%%%%%%%%%%%%%%%%%%%%%%%%%%
\begin{equation}
\tilde{L}_{eff}=2\pi\sum_p(\tilde{c}_i^p \dot{x}_i^p+\tilde{c}^p_{\alpha}\dot{\theta}_{\alpha}^p)-V,
\end{equation}
%%%%%%%%%%%%%%%%%%%%%%%%%%%%%%%%%%%%%%%%%%%%%%%%%%%%%%%%%%%%%%%%%%%%%
where $(x^p_i,\theta^p_{\alpha})$ denotes the position of the $p$-th super vortex,
 and $V$ represents  electric interactions between super vortices.
Since $\tilde{L}_{eff}$  is formally equivalent to the charged particle Lagrangian
 (\ref{oneparticelonsupersphere})
 with $m=0$, % $(B/m \rightarrow \infty)$, 
  the super vortex motion is similar to the particle motion in the LLL.  
The equations of motion for  super vortex are derived as 
%%%%%%%%%%%%%%%%%%%%%%%%%%%%%%%%%%%%%%%%%%%%%%%%%%%%%%%%%%%%%%%%%%%
\begin{subequations}
\begin{align}
&2\pi(-\tilde{f}_{ij}\dot{x}_j+\tilde{f}_{i\alpha}\dot{\theta}_{\alpha})=\mathcal{E}_i,\\
&2\pi(\tilde{f}_{i\alpha}\dot{x}_i+\tilde{f}_{\alpha\beta}\dot{\theta}_{\beta})=
C_{\alpha\beta}\mathcal{E}_{\beta},
\end{align}
\end{subequations}
%%%%%%%%%%%%%%%%%%%%%%%%%%%%%%%%%%%%%%%%%%%%%%%%%%%%%%%%%%%%%%%%%%%
where $\mathcal{E}_i=-{\partial}_i V$ and 
$\mathcal{E}_{\alpha}=C_{\alpha\beta}{\partial}_{\beta}V$.
From these equations, we obtain the super Hall law for vortex 
%%%%%%%%%%%%%%%%%%%%%%%%%%%%%%%%%%%%%%%%%%%%%%%%%%%%%%%%%%%%%%%%%%
\begin{equation}
\mathcal{E}_i \dot{x}_i+C_{\alpha\beta}\mathcal{E}_{\alpha}\dot{\theta}_{\beta}=0.
\end{equation}
%%%%%%%%%%%%%%%%%%%%%%%%%%%%%%%%%%%%%%%%%%%%%%%%%%%%%%%%%%%%%%%%
Thus, the super vortex moves perpendicularly to the direction in  which it is pushed.

%%%%%%%%%%%%%%%%%%%%%%%%%%%%%%%%%%%%%%%%%%%%%%%%%%%%%%%%%%%%%%%%%%%%%%%
\section{summary and discussions}\label{summarysec}
%%%%%%%%%%%%%%%%%%%%%%%%%%%%%%%%%%%%%%%%%%%%%%%%%%%%%%%%%%%%%%%%%%%%%%%%%

We studied  SUSY extensions of the CS theory and the effective field theory
  for the SUSY QHE.
 First, a Lagrangian formalism for the one-particle 
 on the supersphere in the supermonopole background was explored.
The particle motion exhibits a  SUSY generalization of that on the 
 bosonic sphere in the Dirac monopole background, for instance, 
 Hall orthogonality, cyclotron motion and noncommutative geometry.
Next, we constructed a SUSY CS theory with  $OSp(1|2)$
 global supersymmetry.
Our  SUSY CS theory contains many analogous properties 
peculiar to the original CS theory, 
 such as 
 $U(1)$ gauge symmetry,  
 topological mass generation and etc. 
In particular, we derived  a SUSY generalization of the Hopf term, which is 
expressed by highly nonlocal interactions.
Finally, the CSLG description for the SUSY QH liquid was presented.
Based on the duality between the super matter  currents and  the super vector field strengths, 
 we derived the dual CSLG theory, in which 
  super vortices are coupled to  the dual SUSY Maxwell-CS gauge fields. 
It was  shown that 
the super vortex carries the fractional charge  and the vortex motion is equivalent to that  of
 the charged particle in the super LLL.

The SUSY CS theory discussed in this paper is a global $OSp(1|2)$
  extension of  the  simplest CS theory  in  3-dimensional space.
 It would be quite interesting to see 
 SUSY generalizations of the CS theory in higher dimensions.
Their constructions  may be performed  based on the higher dimensional SUSY Lie group, 
 such as $OSp(1|4)$.  
Higher dimensional SUSY CS theories may have  deep connections with supertwistor theory, 
 supermatrix model in higher dimensions, etc.   
Because of the incompressible property of QH liquid, the low energy excitations are confined  
on the edge, and the QH dynamics is well described only by the edge states. 
 We hope to report  edge excitations in SUSY QH liquids 
based on the SUSY CSLG theory in a future work.
The study of the edge excitations  may reveal yet unknown
 dynamical properties of the SUSY QH liquid.
Since, at  present, the SUSY QHE provides a unique physical set-up whose underlying mathematics are
given by the non-anticommutative geometry, it would be quite worthwhile 
 to accomplish full analyses of the SUSY QHE.  
  Such explorations are  expected to  lead to the deeper
 understanding of novel physics in the non-anticommutative world.
 
%%%%%%%%%%%%%%%%%%%%%%%%%%%%%%%%%%%%%%%%%%%%%%%%%%%%%%%%%%%%%%%%%%%
\section*{ACKNOWLEDGEMENTS} 

I would like to acknowledge Satoshi Iso for  valuable comments.
I also thank Masanori Hanada, Masatoshi Sato and Tatsuya Tokunaga  for useful conversations.
Most of this work was performed during the stay at   ISSP and KEK. 
I appreciate the warm hospitalities in the both institutes. 

\appendix

%%%%%%%%%%%%%%%%%%%%%%%%%%%%%%%%%%%%%%%%%%%%%%%%%%%%%%%%%%%%%%%%%
%%%%%%%%%%%%%%%%%%%%%%%%%%%%%%%%%%%%%%%%%%%%%%%%%%%%%%%%%%%%%%%%%

\section{Several formulas about Supermatrix }\label{somformulurasupermatrix}

When the supermatrix $X$ takes  the form of
%%%%%%%%%%%%%%%%%%%%%%%%%%%%%%%%%%%%%%%%%%%%%%%%%
\begin{equation}
X=
\begin{pmatrix}
M & P \\
 Q & N
\end{pmatrix},
\end{equation}
%%%%%%%%%%%%%%%%%%%%%%%%%%%%%%%%%%%%%%%%%%%%%%%%%%
(where $M$ and $N$ are Grassmann even matrices,  $P$ and $Q$ are Grassmann 
 odd matrices)
its superdeterminant is given by 
%%%%%%%%%%%%%%%%%%%%%%%%%%%%%%%%%%%%%%%%%%%%%%%%%%%
\begin{equation}
sdet X= \frac{det(M-PN^{-1}Q)}{det N}=\frac{det M}{det (N-QM^{-1}P)},
\label{superdeterminantformula}
\end{equation}
%%%%%%%%%%%%%%%%%%%%%%%%%%%%%%%%%%%%%%%%%%%%%%%%%%%%
and  the super-adjoint is defined  as 
%%%%%%%%%%%%%%%%%%%%%%%%%%%%%%%%%%%%%%%%%%%%%%%%%%%%%%
\begin{equation}
X^{\ddagger}=
\begin{pmatrix}
 M^{\dagger} & -Q^{\dagger} \\
 P^{\dagger} & N^{\dagger}
\end{pmatrix}.
\label{super-adjointformula}
\end{equation}
%%%%%%%%%%%%%%%%%%%%%%%%%%%%%%%%%%%%%%%%%%%%%%%%%%%%%%%
It is noted that our definition of  super-adjoint is different from the conventional one. 
(In the conventional definition, the minus sign is placed in front of $P^{\dagger}$,  not  $Q^{\dagger}$.) 
Similarly, the supertrace is defined as 
%%%%%%%%%%%%%%%%%%%%%%%%%%%%%%%%%%%%%%%%%%%%%%%%%%%%%%%%%%%%%
\begin{equation}
str(X)=tr(M)-tr(N).
\label{supertraceformula}
\end{equation}
%%%%%%%%%%%%%%%%%%%%%%%%%%%%%%%%%%%%%%%%%%%%%%%%%%%%%%%%%%%%%%
For instance, for the fundamental representations (\ref{fundamentalgene}), the supertraces are  calculated as 
%%%%%%%%%%%%%%%%%%%%%%%%%%%%%%%%%%%%%%%%%%%%%%%%%%%%%%%%%%%%%%
\begin{subequations}
\begin{align}
&str(l_al_b)=\frac{1}{2}\delta_{ab},\\
&str(l_al_{\alpha})=0,\\
&str(l_{\alpha}l_{\beta})=-\frac{1}{2}C_{\alpha\beta}.
\end{align}
\label{supertracefund}
\end{subequations}
%%%%%%%%%%%%%%%%%%%%%%%%%%%%%%%%%%%%%%%%%%%%%%%%%%%%%%%%%%%%%%%%%

The inverse of $X$ is given by 
%%%%%%%%%%%%%%%%%%%%%%%%%%%%%%%%%%%%%%%%%%%%%%%%%%%%%%%%%%%%%%%%%%%%%%%
\begin{align}
&X^{-1}=
\begin{pmatrix}
  \frac{1}{M-P\frac{1}{N}Q}   &   - \frac{1}{M-P\frac{1}{N}Q}  P\frac{1}{N}\\
 -\frac{1}{N}Q\frac{1}{M-P\frac{1}{N}Q}  & \frac{1}{N-Q\frac{1}{M}P}
\end{pmatrix}\nonumber\\
&~~~~~~=
\begin{pmatrix}
 \frac{1}{M-P\frac{1}{N}Q}     &-\frac{1}{M}P\frac{1}{N-Q\frac{1}{M}P}\\
 -\frac{1}{N-Q\frac{1}{M}P}Q\frac{1}{M}    & \frac{1}{N-Q\frac{1}{M}P}
\end{pmatrix}.
\label{inversexformu}
\end{align}
%%%%%%%%%%%%%%%%%%%%%%%%%%%%%%%%%%%%%%%%%%%%%%%%%%%%%%%%%%%%%%%%%%%%%%%%
When $X$ takes the special form
%%%%%%%%%%%%%%%%%%%%%%%%%%%%%%%%%%%%%%%%%%%%%%%%%%%%%%%%%%%%%%%%
\begin{equation}
X=
\begin{pmatrix}
 M &P\\
 -P^t & N
\end{pmatrix},
\label{superxmat}
\end{equation}
%%%%%%%%%%%%%%%%%%%%%%%%%%%%%%%%%%%%%%%%%%%%%%%%%%%%%%%%%%%%%%%%%
its inverse is calculated as 
%%%%%%%%%%%%%%%%%%%%%%%%%%%%%%%%%%%%%%%%%%%%%%%%%%%%%%%%%%%%%%
\begin{equation}
X^{-1}=
\begin{pmatrix}
 \tilde{M} & -\tilde{M}P\frac{1}{N} \\
\frac{1}{N}P^t\tilde{M} & \tilde{N}
\end{pmatrix}
=\begin{pmatrix}
 \tilde{M} & - \frac{1}{M}P\tilde{N} \\
\tilde{N}P^t \frac{1}{M} & \tilde{N}
\end{pmatrix},
\label{xinverse}
\end{equation}
%%%%%%%%%%%%%%%%%%%%%%%%%%%%%%%%%%%%%%%%%%%%%%%%%%%%%%%%%%%%%
where $\tilde{M}$ and $\tilde{N}$ are defined in Eqs.(\ref{tildemn}).

\section{Super Jacobi and Bianchi identities}\label{superjacobisec}

With arbitrary operators $(A,B,C)$ (any of $A,B,C$ can be  Grassmann even or odd), 
it is easy to check  the super Jacobi identities 
%%%%%%%%%%%%%%%%%%%%%%%%%%%%%%%%%%%%%%%%%%%%%%%%%%%%%%%%%%%%%%%%%%%%%%%%%%%%%
\begin{subequations}
\begin{align}
&[[A,B],C]+[[B,C],A]+[[C,A],B]=0, \label{jacobi1} \\
&\{[A,B],C\}+[\{B,C\},A]-\{[C,A],B\}=0,\label{jacobi2} \\
&[\{A,B\},C]+[\{B,C\},A]+[\{C,A\},B]=0\label{jacobi3}.
\end{align}
\end{subequations}
%%%%%%%%%%%%%%%%%%%%%%%%%%%%%%%%%%%%%%%%%%%%%%%%%%%%%%%%%%%%%%%%%%%%%%%%%%%%%%%%
Substituting $(A,B,C)=(D_a,D_b,D_c)$  (where $D_a=\partial_a+iA_a$) and 
$(A,B,C)=(D_a,D_b,D_{\alpha})$  (where $D_{\alpha}=\partial_{\alpha}+iA_{\alpha}$)
 to the 1st Jacobi identity (\ref{jacobi1}) respectively, we 
obtain the 1st 
Bianchi identity
%%%%%%%%%%%%%%%%%%%%%%%%%%%%%%%%%%%%%%%%%%%%%%%%%%%%%%%%%%%%%%%%%%%%%%%%%%%%
\begin{equation}
\partial_a F_{bc}+\partial_b F_{ca}+\partial_{c}F_{ab}=0, \label{superb1}
\end{equation}
%%%%%%%%%%%%%%%%%%%%%%%%%%%%%%%%%%%%%%%%%%%%%%%%%%%%%%%%%%%%%%%%%%%%%%%%%%%%%%
and 
 the 2nd Bianchi identity  
%%%%%%%%%%%%%%%%%%%%%%%%%%%%%%%%%%%%%%%%%%%%%%%%%%%%%%%%%%%%%%%%%%%%%%%%%%%%%
\begin{equation}
\partial_a F_{b\alpha}+\partial_b F_{\alpha a}+\partial_{\alpha}F_{ab}=0.\label{superb2}
\end{equation}
%%%%%%%%%%%%%%%%%%%%%%%%%%%%%%%%%%%%%%%%%%%%%%%%%%%%%%%%%%%%%%%%%%%%%%%%%%%%%
Similarly, substituting $(A,B,C)=(D_a,D_{\alpha},D_{\beta})$ to 
 the 2nd Jacobi identity (\ref{jacobi2}), the 3rd
Bianchi identity 
%%%%%%%%%%%%%%%%%%%%%%%%%%%%%%%%%%%%%%%%%%%%%%%%%%%%%%%%%%%%%%%%%%%%%%%%%%%%
\begin{equation}
\partial_a F_{\alpha\beta}-\partial_{\alpha} F_{a\beta}-\partial_{\beta}F_{a\alpha}=0
\label{superb3}
\end{equation}
%%%%%%%%%%%%%%%%%%%%%%%%%%%%%%%%%%%%%%%%%%%%%%%%%%%%%%%%%%%%%%%%%%%%%%%%%%%%%%
is obtained.
 From the last
 identity (\ref{jacobi3}), with $(A,B,C)=(D_{\alpha},D_{\beta},D_{\gamma})$,  we obtain the 
last Bianchi identity  
%%%%%%%%%%%%%%%%%%%%%%%%%%%%%%%%%%%%%%%%%%%%%%%%%%%%%%%%%%%%%%%%%%%%%%%%%%%%%
\begin{equation}
\partial_{\alpha} F_{\beta\gamma}+\partial_{\beta} F_{\gamma\alpha}+\partial_{\gamma}
F_{\alpha\beta}=0. \label{superb4}
\end{equation}
%%%%%%%%%%%%%%%%%%%%%%%%%%%%%%%%%%%%%%%%%%%%%%%%%%%%%%%%%%%%%%%%%%%%%%%%%%%%%


\begin{thebibliography}{99}

\bibitem{cond-mat/0110572}
Shou-Cheng Zhang, Jiangping Hu,
{\it ``A Four Dimensional Generalization of the Quantum Hall Effect''},
Science 294 (2001) 823; cond-mat/0110572. 

\bibitem{cond-mat/0306045}
 B.A. Bernevig, J.P. Hu, N. Toumbas, S.C. Zhang,
{\it ``The Eight Dimensional Quantum Hall Effect and the Octonions''},
Phys.Rev.Lett. 91 (2003) 236803; cond-mat/0306045.


\bibitem{hep-th/0310274} 
 Kazuki Hasebe, Yusuke Kimura, 
{\it ``Dimensional Hierarchy in Quantum Hall Effects on Fuzzy Spheres''},
 Phys.Lett.B602(2004) 255-260; hep-th/0310274.


\bibitem{hep-th/0203264}
 Dimitra Karabali, V.P. Nair,
{\it ``Quantum Hall Effect in Higher Dimensions''},  
Nucl.Phys. B641 (2002) 533-546; hep-th/0203264.


\bibitem{hep-th/0309212}
 V.P. Nair, S. Randjbar-Daemi, 
{\it ``Quantum Hall effect on $S^3$, edge states and fuzzy $S^3/{\bf Z}_2$''},
Nucl.Phys. B679 (2004) 447-463; hep-th/0309212.


\bibitem{hep-th/0505095} 
Ahmed Jellal,
{\it ``Quantum Hall Effect on Higher Dimensional Spaces''},
Nucl.Phys.B725 (2005) 554-576; hep-th/0505095. 



\bibitem{hep-th/0504092} 
Giovanni Landi,
{\it``Spin-Hall effect with quantum group symmetry''},
 Lett.Math.Phys.75 (2006) 187-200; hep-th/0504092. 





\bibitem{hep-th/0302109}
Hirosi Ooguri, Cumrun Vafa,
{\it ``The C-Deformation of Gluino and Non-planar Diagrams''},
Adv.Theor.Math.Phys. 7 (2003) 53-85, hep-th/0302109; 
%\bibitem{hep-th/0303063}
% Hirosi Ooguri, Cumrun Vafa,
{\it ``Gravity Induced C-Deformation''},
Adv.Theor.Math.Phys. 7 (2004) 405-417, hep-th/0303063. 

\bibitem{hep-th/0302078}
J. de Boer, P. A. Grassi, P. van Nieuwenhuizen,
{\it ``Non-commutative superspace from string theory''},
Phys.Lett. B574 (2003) 98-104, hep-th/0302078. 

\bibitem{hep-th/0305248}
Nathan Seiberg,
{\it ``Noncommutative Superspace, N=1/2 Supersymmetry, Field Theory and String Theory''},
JHEP 0306 (2003) 010, hep-th/0305248. 


\bibitem{hep-th/0411137}
 Kazuki Hasebe,
{\it ``Supersymmetric Quantum Hall Effect on a Fuzzy Supersphere''},
Phys.Rev.Lett. 94 (2005) 206802; hep-th/0411137. 



\bibitem{hep-th/0503162}
 Kazuki Hasebe,
{\it ``Quantum Hall Liquid on a Noncommutative Superplane''},
Phys.Rev.D72 (2005) 105017; hep-th/0503162. 


\bibitem{hep-th/0311159}
 Evgeny Ivanov, Luca Mezincescu, Paul K. Townsend,
{\it ``Fuzzy $CP(n|m)$ as a quantum superspace''},
hep-th/0311159, {\it``A Super-Flag Landau Model''},
hep-th/0404108, {\it``Planar Super-Landau Models''}, 
JHEP 0601 (2006) 143; hep-th/0510019.



\bibitem{hep-th/0503244}
 S. Bellucci, A. Beylin, S. Krivonos, A. Nersessian, E. Orazi,
{\it ``N=4 supersymmetric mechanics with nonlinear chiral supermultiplet''},
Phys.Lett. B616 (2005) 228-232; hep-th/0503244.


\bibitem{hep-th/0410070}
James Gates Jr, Ahmed Jellal, EL Hassan Saidi, Michael Schreiber,
{\it ``Supersymmetric Embedding of the Quantum Hall Matrix Model''},
JHEP 0411 (2004) 075; hep-th/0410070.

\bibitem{PRL8862}
 S.C. Zhang, T.H. Hansson, S. Kivelson
{\it ``An Effective Field Theory Model For The Fractional Quantum Hall Effect''},
Phys.Rev.Lett.62 (1989) 82-85. 

\bibitem{IJMP92B6}
S.C. Zhang,
{\it ``The Chern-Simons-Landau-Ginzburg Theory Of The Fractional Quantum Hall Effect''}, 
 Int.J.Mod.Phys. B6 (1992) 25-58.

\bibitem{cond-mat/0206164}
 B.A. Bernevig, C.H. Chern, J.P. Hu, N. Toumbas, S.C. Zhang,
{\it ``Effective field theory description of the higher dimensional quantum Hall liquid''},
Annals Phys. 300 (2002) 185; cond-mat/0206164 



\bibitem{hep-th/9902115}
   See, for instance, Gerald V. Dunne,
{\it ``Aspects of Chern-Simons Theory''},
  hep-th/9902115. 



\bibitem{math-ph/9907020}
 Giovanni Landi,
{\it ``Projective Modules of Finite Type over the Supersphere $S^{2,2}$''},
Differ.Geom.Appl. 14 (2001) 95-111; math-ph/9907020. 

\bibitem{hep-th/0409230}
Kazuki Hasebe, Yusuke Kimura, 
{\it ``Fuzzy Supersphere and Supermonopole''},
Nucl.Phys. B709 (2005) 94-114; hep-th/0409230. 






\bibitem{hep-th/9612031}
Kiyoshi Ezawa, Atushi Ishikawa,
{\it ``$Osp(1|2)$ Chern-Simons gauge theory as 2D N=1 Induced Supergravity''},
Phys.Rev. D56 (1997) 2362-2368; hep-th/9612031.


\bibitem{PhysRevD45}
Bum-Hoon Lee, Choonkyu Lee and Hyunsoo Min,
{\it ``Supersymmetric Chern-Simons vortex systems and fermion zero modes''},
Phys. Rev. D 45 (1992) 4588-4599. 


\bibitem{hep-th/0311005}
 Satoshi Iso, Hiroshi Umetsu,
{\it ``Gauge Theory on Noncommutative Supersphere from Supermatrix Model''},
Phys.Rev. D69 (2004) 105003; hep-th/0311005. 


\bibitem{PhysRevD41}
G. V. Dunne, R. Jackiw, and C. A. Trugenberger, 
{\it ``Topological (Chern-Simons) quantum mechanics''},
Phys. Rev. D 41 (1990) 661-666. 



\bibitem{hep-th/9202012}
  Satoshi Iso, Dimitra Karabali, B. Sakita,
{\it ``One-Dimensional Fermions as Two-Dimensional Droplets Via Chern-Simons Theory ''},
 Nucl.Phys. B388 (1992) 700-714; hep-th/9202012. 


\end{thebibliography}
\end{document}